\documentclass[pre,aps,twocolumn,superscriptaddress]{revtex4-1}

\usepackage{graphicx}
\usepackage{amssymb,amsfonts,amsmath}
\usepackage{color}
\usepackage{ulem}
\usepackage[hidelinks]{hyperref}
\usepackage{bbm}
\usepackage{bbold}
\usepackage{algpseudocode}
\usepackage{algorithm}

\usepackage{tikz}
\usetikzlibrary{shapes}
\usetikzlibrary{patterns}
\usetikzlibrary{angles, quotes}

\newcommand{\rv}{{\mathbf r}}
\newcommand{\Rv}{{\mathbf R}}

\newcommand{\Tr}{{\rm Tr}\,}
\newcommand{\e}{{\rm e}}

\newcommand{\Jv}{{\bf J}}

\newcommand{\pv}{{\bf p}}

\newcommand{\Pv}{{\bf P}}

\newcommand{\Fv}{{\bf F}}
\newcommand{\fv}{{\bf f}}

\newcommand{\msphantom}[1]{$\ldots$}

\newcommand{\eps}{{\boldsymbol \epsilon}}
\newcommand{\nablai}{\nabla_i}

\newcommand{\eqr}[1]{Eq.~\eqref{#1}}

\newcommand{\unity}{{\mathbbm 1}}

\newcommand{\mydelete}[1]{{}}
\newcommand{\taub}{{\boldsymbol\tau}}

\newcommand{\rmint}{{\rm int}}

\newcommand{\rmext}{{\rm ext}}

\newcommand{\bsig}{\boldsymbol\sigma}
\newcommand{\Sv}{{\bf S}}

\newcommand{\sv}{{\mathbf r}}
\newcommand{\rvi}{\rv_i}
\newcommand{\pvi}{\pv_i}

\begin{document}

\title{Why gauge invariance applies to statistical mechanics}

\author{Johanna M\"uller}
\affiliation{Theoretische Physik II, Physikalisches Institut, 
  Universit{\"a}t Bayreuth, D-95447 Bayreuth, Germany}

\author{Florian Samm\"uller}
\affiliation{Theoretische Physik II, Physikalisches Institut, 
  Universit{\"a}t Bayreuth, D-95447 Bayreuth, Germany}

\author{Matthias Schmidt}
\affiliation{Theoretische Physik II, Physikalisches Institut, 
  Universit{\"a}t Bayreuth, D-95447 Bayreuth, Germany}
\email{Matthias.Schmidt@uni-bayreuth.de}

\date{8 October 2024, 
  second revision: 6 March 2025}
\begin{abstract}
  We give an introductory account of the recently identified gauge
  invariance of the equilibrium statistical mechanics of classical
  many-body systems [J.~M\"uller {\it et
      al.},~\href{https://doi.org/10.1103/PhysRevLett.133.217101}
    {Phys. Rev. Lett. {\bf 133}, 217101 (2024)}]. The
  gauge transformation is a non-commutative shifting operation on
  phase space that keeps the differential phase space volume element
  and hence the Gibbs integration measure conserved. When thermally
  averaged any observable is an invariant, including thermodynamic and
  structural quantities. Shifting transformations are canonical in the
  sense of classical mechanics. They also form an infinite-dimensional
  group with generators of infinitesimal transformations that build a
  non-commutative Lie algebra. We lay out the connections with the
  underlying geometry of coordinate displacement and with Noether's
  theorem.  Spatial localization of the shifting yields differential
  operators that satisfy commutator relationships, which we describe
  both in purely configurational and in full phase space
  setups. Standard operator calculus yields corresponding equilibrium
  hyperforce correlation sum rules for general observables and order
  parameters.  Using Monte Carlos simulations we demonstrate
  explicitly the gauge invariance for finite shifting. We argue in
  favour of using the gauge invariance as a statistical mechanical
  construction principle for obtaining exact results and for
  formulating smart sampling algorithms.
\end{abstract}

\maketitle

\section{Introduction}
\label{SECintroduction}

Gauge field theories are fundamental for our understanding of
nature. Considering the invariance against well-defined gauge
transformations constitutes a universal construction principle, which
formalizes the independence of the physical predictions that a theory
provides from the precise choice of the variables that are in use. The
perhaps most well-known case of gauge invariance affects the (scalar
and vector) potentials of electrodynamics. While thereby specific
choices of gauge can significantly simplify practical calculations,
arguably as important are the fundamental consequences of gauge
invariance.  In the present example it is the intimate connection with
the local conservation of electrical charge. We describe this standard
case below as a template for the many-body physics that we address in
this contribution.

Noether's theorem of invariant variations \cite{noether1918,
  byers1998, brading2002, read2022book} provides the appropriate
mathematical framework for systematically addressing the consequences
that follow from the juxtaposition of inherent independence and
apparent dependence on the choice of gauge. While forming a staple of
field theories for many decades, Noether's theorem~\cite{noether1918,
  byers1998, brading2002, read2022book} has seen only relatively
recently an increasing number of applications to statistical physics
\cite{revzen1970, baez2013markov, marvian2014quantum, sasa2016,
  sasa2019, bravetti2023, budkov2022, brandyshev2023, budkov2024ionic,
  budkov2024jcp, beyen2024, beyen2024generic}, both in and
out-of-equilibium.  As much of statistical physics is based on
insightful combination of suitable approximations with exact results,
perhaps most notably in the form of equilibrium sum rules
\cite{hansen2013, evans1979,evans1992, baus1984, evans1990,
  henderson1992, triezenberg1972, percus1962, gul2024testParticle},
there is arguably much potential for making progress.

Applications of the Noether theorem were performed in a variety of
statistical mechanical settings as follows.  In an early study of
functional integrals in statistical physics and primarily based on
quantum field theory, Revzen derived the continuity equation from
Noether's theorem \cite{revzen1970}.  Baez and Fong formulated a
Noether theorem for Markov processes \cite{baez2013markov}.  Marvian
and Spekkens provided an extension of Noether’s theorem by quantifying
the asymmetry of quantum states \cite{marvian2014quantum}.  Sasa and
Yokokura formulated the thermodynamic entropy as a Noether invariant
\cite{sasa2016}. Their work lead to investigation of the
thermodynamical path integral and emergent symmetry \cite{sasa2019}.
Recently the thermodynamic entropy was viewed as a Noether invariant
from contact geometry \cite{bravetti2023}.  Budkov and collaborators
carried out a range of studies of complex systems, where the use of
Noether's theorem was crucial. This work addressed modified
Poisson-Boltzmann equations and macroscopic forces in inhomogeneous
ionic fluids \cite{budkov2022}, as well as a covariant field theory of
mechanical stresses in inhomogeneous ionic fluids
\cite{brandyshev2023}.  The thermomechanical approach was applied to
calculating mechanical stresses in inhomogeneous ionic fluids
\cite{budkov2024ionic} and the surface tension of aqueous electrolyte
solutions \cite{budkov2024jcp}.  In recent work, Beyen and Maes
identified the entropy as a Noether charge for quasistatic gradient
flow \cite{beyen2024} and for more general
setups~\cite{beyen2024generic}.

The statistical mechanics of particle-based systems is well accessible
via direct simulation \cite{frenkel2023book}. Thereby inherent
statistical uncertainties can be reduced considerably by a range of
recent methods, most prominently via mapped averaging
\cite{moustafa2015, schultz2016, moustafa2017jctp, moustafa2017prb,
  schultz2018, purohit2018, moustafa2019, purohit2020, moustafa2022,
  lin2018, trokhymchuk2019, schultz2019, purohit2019} and force
sampling \cite{frenkel2023book, borgis2013,
  delasheras2018forceSampling, coles2019, coles2021, rotenberg2020,
  mangaud2020, coles2023revelsMD,renner2023torqueSampling}.  Mapped
averaging, as pioneered by Kofke and coworkers \cite{moustafa2015,
  schultz2016, moustafa2017jctp, moustafa2017prb, schultz2018,
  purohit2018, moustafa2019, purohit2020, moustafa2022, lin2018,
  trokhymchuk2019, schultz2019, purohit2019}, is a systematic means
for the reformulation of ensemble averages via flexible (coordinate)
mappings \cite{schultz2016}. The approach was used in a variety of
contexts, including sampling of thermal properties of crystals
\cite{moustafa2015, schultz2016, moustafa2017jctp,
  moustafa2017prb,schultz2018, purohit2018, moustafa2019, purohit2020,
  moustafa2022}, of liquid water \cite{lin2018}, and of hard sphere
\cite{trokhymchuk2019} and Lennard-Jones fluids \cite{schultz2018}.
Force sampling originates from a pioneering early investigation by
Borgis {\it et al.}~\cite{borgis2013} and we refer to the review by
Rotenberg \cite{rotenberg2020} for a comprehensive account of the
approach. As a central concept, force sampling is based on using
numerical integration methods to invert spatial gradients of
correlation functions with the aim to exploit the accompanying
smoothening effect.  The relationship of force-sampling methods and
mapped averaging has been illuminated in Ref.~\cite{purohit2019}.

A specific shifting operation on the phase space associated with
general interacting many-body systems was recently put forward and
shown to be useful in a broad range of statistical mechanical settings
\cite{hermann2021noether, hermann2022topicalReview,
  hermann2022variance, hermann2022quantum, sammueller2023whatIsLiquid,
  hermann2023whatIsLiquid, robitschko2024any}.  These applications
include formulating force-based density functional theory
\cite{tschopp2022forceDFT, sammueller2022forceDFT}, self-consistency
conditions for neural functionals \cite{sammueller2023neural,
  sammueller2023whyNeural}, the formulation of hyperdensity functional
theory for the description of the equilibrium behaviour of general
observables \cite{sammueller2024hyperDFT, sammueller2024whyhyperDFT},
as well as nonequilibrium power functional \cite{schmidt2022rmp} sum
rules \cite{hermann2021noether, delasheras2023perspective,
  zimmermann2024ml}.
We refer the Reader to Refs.~\cite{hermann2021noether,
  robitschko2024any} for discussions of the relationship to the body
of existing sum rules in the liquid state literature.

The phase space shifting \cite{hermann2021noether,
  hermann2022topicalReview, hermann2022variance, hermann2022quantum,
  sammueller2023whatIsLiquid, hermann2023whatIsLiquid,
  robitschko2024any, tschopp2022forceDFT, sammueller2022forceDFT} was
recently argued to constitute a local gauge transformation of
statistical mechanical microstates \cite{mueller2024gauge}; see the
recent accounts given by Rotenberg \cite{rotenberg2024spotted} and by
Miller \cite{miller2025physicsToday}. Here we provide in-depth
background for this gauge shifting
transformation~\cite{mueller2024gauge}. The transformation is
geometric in nature and we describe in detail its origins in spatial
displacements of particle coordinates. We introduce a purely
configurational version and show that including the complementary
momentum transformation \cite{mueller2024gauge}, realized by
(position-resolved) matrix multiplication, completes a very specific
mathematical structure. As its key features, the shifting
transformation is both canonical in the sense of classical mechanics
and it constitutes a non-commutative continuous group. Phase space
shifting leaves any equilibrium average invariant, which establishes
its status as a gauge transformation. Meeting expectations for this
role, the shifting unambiguously leads to a breadth of exact
identities.

Our gauge theory carries strong similarities in its mathematical
structure with the theory of Lie groups and corresponding Lie algebras
of infinitesimal transformations.  Important examples of nontrivial
Lie groups in physics include the rotation matrices of
three-dimensional space and the Pauli matrices to describe quantum
mechanical spin 1/2 particles \cite{goldstein2002}.  Our theory goes
in important ways further than these elementary cases. In particular
the tight interweaving with spatial Dirac distributions is
notable. This feature provides microscopically sharp resolution and,
as we demonstrate, it allows one to retain particle-level precision
for all ensuing force and hyperforce correlation functions and their
sum rules.

The paper is organized as follows. We describe in
Sec.~\ref{SECelectrodynamics} the standard gauge invariance of
classical electrodynamics to both provide a discussion of general
concepts of gauge invariance and also to formulate a blueprint for the
subsequent particle-based physics.  Readers who are familiar with this
material are welcome to skip forward to
Sec.~\ref{SECgaugeInvarianceOnPhaseSpace} where we describe the
statistical mechanical phase space gauge invariance. The underlying
phase space shifting transform being presented in
Sec.~\ref{SECfiniteShifting}. Its central properties are derived in
Sec.~\ref{SECfinitePhaseSpaceShifting} and the transform is thereby
taken to be in general of finite magnitude.  Rich additional
mathematical structure is revealed when considering infinitesimal
transforms as is described in detail for particle configuration
(position) shifting in Sec.~\ref{SECinfinitesimalPositionShifting}.

We turn to statistical mechanics in Sec.~\ref{SECstatisticalMechanics}
and first define the general concepts, including the partition sum and
thermal averages in Sec.~\ref{SECpartitionSumAndThermalAverages}. The
relevant standard one-body observables are introduced in
Sec.~\ref{SEConeBodyObservables}.  Position shifting is applied to
derive exact sum rules in Sec.~\ref{SECconfigurationShifting}. General
phase space shifting, which includes additional momentum displacement
and restores the full canonical phase space transform, is described in
Sec.~\ref{SECphaseSpaceShiftingSumRules}.  We present implications for
and applications to computer simulation work in
Sec.~\ref{SECsimulations}.  We give our conclusions and an outlook on
possible future work in Sec.~\ref{SECconclusions}.

\begin{figure}[!t]
  \vspace{1mm}
  \includegraphics[width=.99\columnwidth]{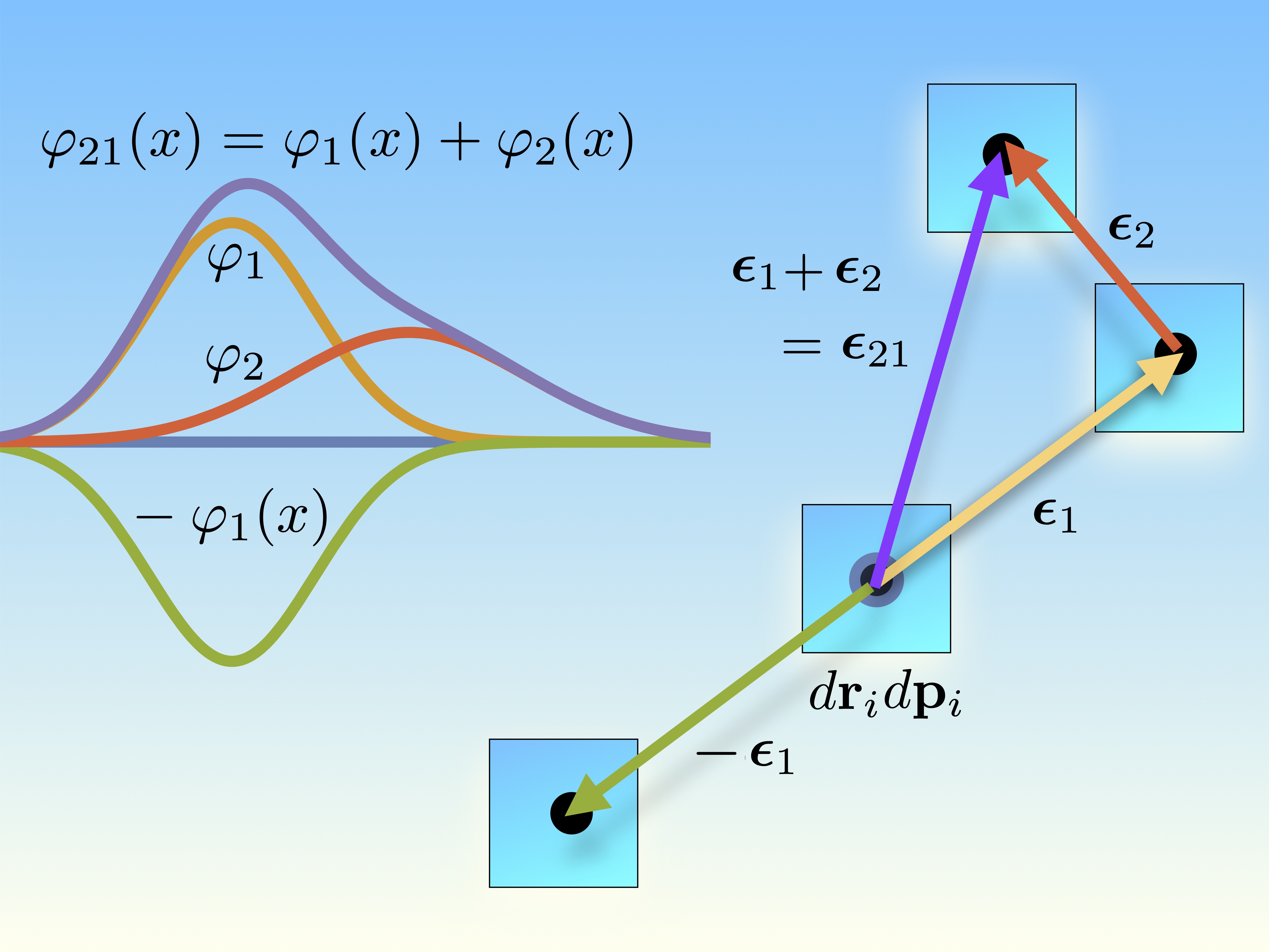}
  \caption{Group structure of the gauge transformations in
    electrodynamics (curves) and in statistical mechanics (arrows),
    depicted for global transforms for the latter case. Shown are
    different forms of the scalar gauge function $\varphi(x)$ that
    parameterizes the transform in electrodynamics. A given gauge
    function $\varphi_1(x)$ has an inverse $-\varphi_1(x)$. Two
    transforms $\varphi_1(x)$ and $\varphi_2(x)$ combine additively to
    $\varphi_{21}(x)=\varphi_1(x)+\varphi_2(x)$.  For the statistical
    mechanics the global gauge invariance is shown for different
    realizations of a constant shifting vector $\eps_1$. The inverse
    transform $\eps_{-1}$ is represented by the negative shifting
    vector, $\eps_{-1}=-\eps_1$. Two transformations that correspond
    to $\eps_1$ and $\eps_2$ combine additively to
    $\eps_{21}=\eps_1+\eps_2$. Also illustrated is the conserved phase
    space volume element $d\rvi d\pvi$.  }
\label{FIGsketchGlobal}
\end{figure}

\section{Classical electrodynamics as a precursor}
\label{SECelectrodynamics}

The prototypical and arguably most well-known example of gauge
invariance in physics is that of classical electrodynamics
\cite{kobe1980, brading2002, iqbal2024}, where the position- and
time-dependent scalar electric potential $V(\rv,t)$ and the magnetic
vector potential ${\bf A}(\rv,t)$ are transformed on the basis of a
scalar gauge function $\varphi(\rv,t)$. The potentials are transformed
according to the map:
\begin{align}
   V(\rv,t) &\to V(\rv,t) -  \partial \varphi(\rv,t)/\partial t,
   \label{EQscalarPotentialTransform}\\
   {\bf A}(\rv,t) &\to {\bf A}(\rv,t) + \nabla \varphi(\rv,t),
   \label{EQvectorPotentialTransform}
\end{align}
where $\nabla$ indicates the derivative with respect to
position~$\rv$.

The form of the gauge function $\varphi(\rv,t)$ can be arbitrarily
chosen, as it has no discernible effect on Maxwell's equations, which
govern the dynamics of the magnetic induction ${\bf
  B}(\rv,t)=\nabla\times {\bf A}(\rv,t)$ and of the electrical field
${\bf E}(\rv,t)=-\nabla V(\rv,t) - \partial {\bf A}(\rv,t)/\partial
t$. This gauge invariance can readily be verified.  That the
transformation preserves the form of ${\bf B}(\rv,t)$ follows from
elementary vector calculus via $\nabla\times \nabla
\varphi(\rv,t)=0$. Likewise ${\bf E}(\rv,t)$ remains unchanged as the
mixed partial derivative of $\varphi(\rv,t)$ with respect to $\rv$ and
$t$ can be interchanged and hence the two terms that emerge from the
gauge transformation cancel each other such that $\nabla \partial
\varphi(\rv,t)/\partial t- \partial \nabla \varphi(\rv,t)/\partial
t=0$.

The gauge freedom of choice of the form of $\varphi(\rv,t)$ has very
profound implications quantum mechanically and a rich structure of
modern theoretical physics is based on different forms of gauge
symmetry. As background here we lay out the gauge freedom for the
classical electrodynamical case; Ref.~\cite{stokes2022} reviews gauge
freedom in quantum electrodynamics.

We use the standard covariant formulation with the four-vector
potential $A_\nu(x) = (V(\rv,t)/c, {\bf A}(\rv,t))$ where the
four-vector $x_\nu=(ct,\rv)$ indicates the spacetime point, $c$ is the
speed of light, and $\nu$ denotes the component of spacetime.  The
electrical charge density $q(\rv,t)$ and the electrical current
density $\Jv(\rv,t)$ form a four-current $J_\nu(x)=(cq(\rv,t),
\Jv(\rv,t))$.  The electromagenetic field tensor then follows as
$F_{\nu\lambda}(x)= \partial_\nu A_\lambda(x) - \partial_\lambda
A_\nu(x)$, where $\partial_\nu$ indicates the derivative with respect
to $x^\nu$, which in explicit form is
$\partial_\nu=(c^{-1}\partial/\partial t, -\nabla)$. All Greek indices
range from 0 (time-like) to 1, 2, and 3 (all space-like) and the
metric has signature $(+,-,-,-)$.

This covariant formalism allows one to express the gauge
transformation equations \eqref{EQscalarPotentialTransform} and
\eqref{EQvectorPotentialTransform} in the succinct form
\begin{align}
  A_\nu(x) \to A_\nu(x) + \partial_\nu \varphi(x).
  \label{EQgaugeTransformationCovariant}
\end{align}

It is revealing to address the consequences of
\eqr{EQgaugeTransformationCovariant} on the standard electromagnetic
action $S$ which consists of contributions that arise from free fields
($S_{\rm free}$) and external influence ($S_\rmext$) according to the
sum $S=S_{\rm free} + S_{\rm ext}$. The two individual terms are given
by
\begin{align}
  S_{\rm free} &= -\frac{1}{4\mu_0}\int dx 
  F_{\lambda\nu}(x) F^{\nu\lambda}(x),
  \label{EQSfree}\\
  S_\rmext &= -\int dx  A_\nu(x) J^\nu(x),
  \label{EQSext}
\end{align}
where $\mu_0$ is the magnetic permeability of the vacuum and we use
Einstein summation convention over pairs of Greek spacetime
indices. Requiring stationarity of $S$ against changes in $A_\nu(x)$
yields upon multiplying by $\mu_0$ the Maxwell equations in the
compact covariant form
\begin{align}
  \partial_\nu F^{\nu\lambda}(x) &= \mu_0 J^\lambda(x).
\end{align}

The free field action $S_{\rm free}$ is already an invariant and we
hence investigate the effects of the gauge transformation on the
external action $S_\rmext$. We apply the transform
\eqref{EQgaugeTransformationCovariant} to \eqr{EQSext} which yields
\begin{align}
  S_\rmext &\to -\int dx [A_\nu(x) + \partial_\nu \varphi(x)]
  J^\nu(x)
  \label{EQSextTransformedFirstStep}
  \\
  &\quad = S_\rmext +\int dx \varphi(x) \partial_\nu J^\nu(x).
  \label{EQSextTransformed}
\end{align}
In \eqr{EQSextTransformedFirstStep} we have identified the first term
in the integral as $S_\rmext$ via \eqr{EQSext} and in the second term
we have integrated by parts and assumed that boundary terms vanish.

Imposing that $S_\rmext$ shall be gauge invariant implies that the
second term on the right hand side of \eqr{EQSextTransformed} must
vanish. Given that the form of the gauge function $\varphi(x)$ is
arbitrary, this can only be true provided that
\begin{align}
  \partial_\nu J^\nu(x) &= 0,
  \label{EQcontinuity}
\end{align}
which is the continuity equation for the charge distribution. When
returning to three-vectors \eqr{EQcontinuity} attains the familiar
more elementary form $\partial q(\rv,t)/\partial t=-\nabla\cdot
\Jv(\rv,t)$.

The simplicity of the specific form \eqref{EQSext} of the external
action allowed us to derive \eqr{EQcontinuity} directly. Instead of
relying explicitly on the fundamental lemma of variational calculus,
we can equivalently utilize functional differentiation, here with
respect to the gauge function $\varphi(x)$.  We lay out these
connections in appendix \ref{SECappendixElectrodynamics}, where we
also demonstrate the gauge invariance of $S_{\rm free}$ explicitly.

Gauge transformations encompass specific group structure that is
associated with repeated application of the transformation. In
particular, chaining together two subsequent transforms constitutes
again a single transform. This is easy to see when considering two
transforms represented by two different gauge functions $\varphi_1(x)$
and $\varphi_2(x)$. Then the composite transform is simply
characterized by the sum $\varphi_{21}(x) = \varphi_1(x) +
\varphi_2(x)$. Clearly, the order of the first and the second
transformations is irrelevant,
$\varphi_{21}(x)=\varphi_{12}(x)=\varphi_2(x)+\varphi_1(x)$, and hence
the group is commutative (Abelian). For completeness, a transform with
given $\varphi_1(x)$ has an inverse transform characterized by
$-\varphi_1(x)$, the neutral group element is $\varphi(x)=0$, and the
composition is associative. These properties establish formally the
mathematical group structure.

An illustration of the local gauge transformation of the potentials of
electrodynamics is shown in Fig.~\ref{FIGsketchGlobal}. We also show a
depiction of global shifts that displace the coordinates $\rv_i$ of
particle $i$ in a many-body system. Statistical mechanical averages
were shown to be invariant under such transformations of all particles
\cite{hermann2021noether, hermann2022topicalReview,
  hermann2022variance} and we refer the reader to
Refs.~\cite{brandyshev2023,brandyshev2024} for similar applications in
field theory. In the following we turn to the mathematically much
richer local version of phase space shifting \cite{hermann2022quantum,
  sammueller2023whatIsLiquid, hermann2023whatIsLiquid,
  robitschko2024any, tschopp2022forceDFT, sammueller2022forceDFT,
  mueller2024gauge}.

\section{Gauge invariance of phase space }
\label{SECgaugeInvarianceOnPhaseSpace}

\subsection{Local shifting as a canonical transformation}
\label{SECfiniteShifting}

As laid out in the introduction, Noether's theorem \cite{noether1918,
  byers1998, brading2002, read2022book} has been applied in
statistical physics in a variety of settings \cite{revzen1970,
  baez2013markov, marvian2014quantum, sasa2016, sasa2019,
  bravetti2023, budkov2022, brandyshev2023, budkov2024ionic,
  budkov2024jcp, beyen2024, beyen2024generic}. Here we describe the
phase space shifting transformation \cite{hermann2021noether,
  hermann2022topicalReview, hermann2022variance, tschopp2022forceDFT,
  hermann2022quantum, sammueller2023whatIsLiquid,
  hermann2023whatIsLiquid, robitschko2024any} that affects the
positions $\rv_i$ and the momenta $\pv_i$ of a classical many-body
system, where the index $i=1,\ldots, N$ enumerates the $N$
particles. The shifting transform \cite{hermann2022quantum} is
parameterized by a smooth (infinitely differentiable)
three-dimensional vector field $\eps(\rv)$, where $\rv$ denotes
spatial position.

The position and momentum degrees of freedom of each particle $i$ are
affected in the same way. Specifically, the phase space transformation
is
\begin{align}
  \rv_i &\to \rv_i + \eps(\rv_i) = \rv_i',
  \label{EQriTransform}\\
  \pv_i &\to [\unity + \nabla_i\eps(\rv_i)]^{-1}\cdot \pv_i = \pv_i',
  \label{EQpiTransform}
\end{align}
where $\unity$ denotes the $d\times d$-unit matrix, with $d$
indicating spatial dimensionality, $\nabla_i$ denotes the derivative
with respect to particle position $\rv_i$, the superscript $-1$
denotes matrix inversion, and the transformed variables are indicated
by a prime (rather than the tilde used in
Refs.~\cite{mueller2024gauge, hermann2023whatIsLiquid,
  robitschko2024any}).  We adopt the convention that the
$ab$-component of $\nabla_i \eps(\rv_i)$ is $[\nabla_i
  \eps(\rv_i)]_{ab}= \nabla_{i,a} \epsilon_{b}(\rv_i)$, where the
indices $a$,$b$ denote the Cartesian components.

\begin{figure}[!t]
  \vspace{1mm}
  \includegraphics[width=.99\columnwidth]{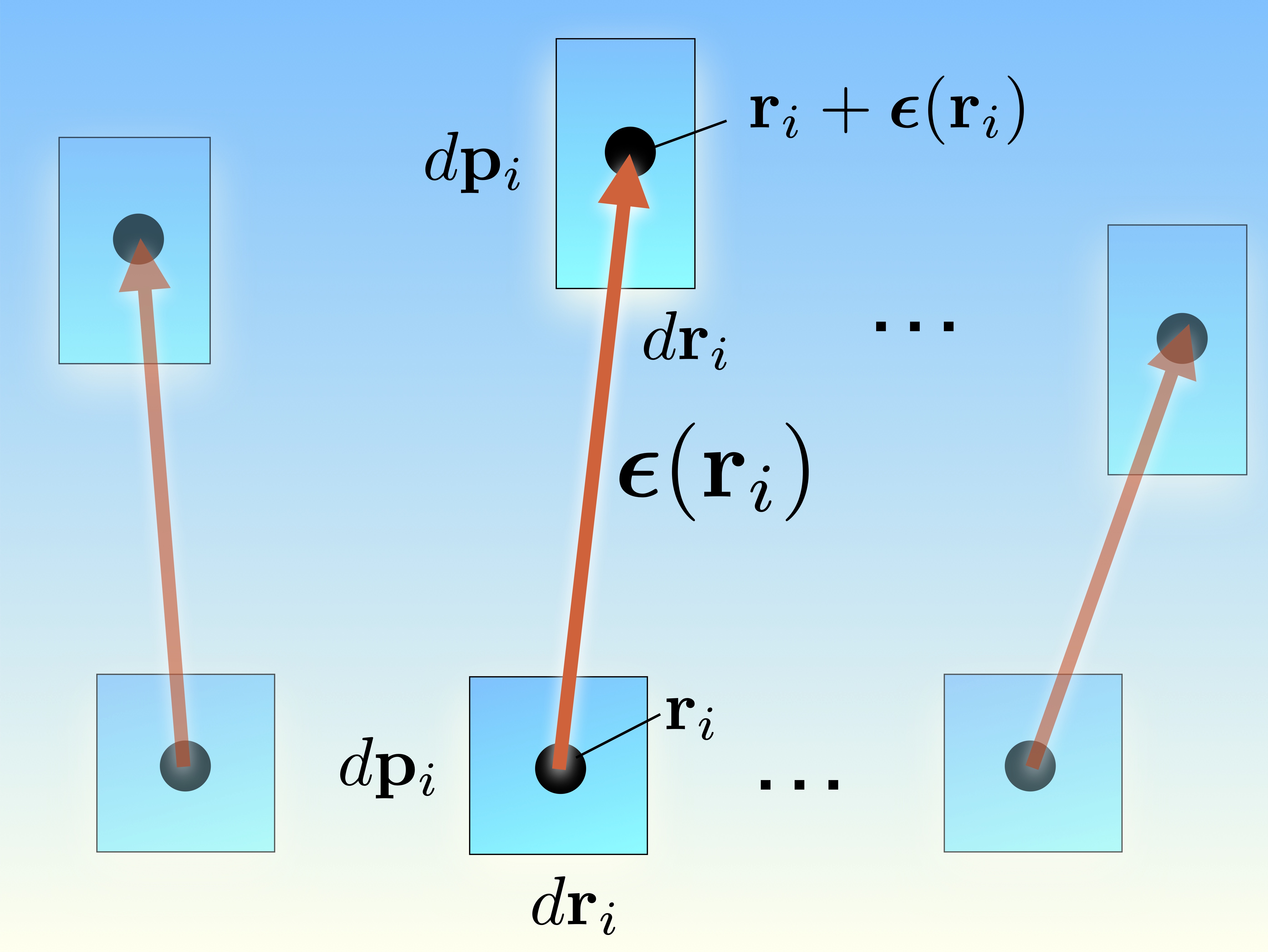}
  \caption{Shifting of microstates. The position and momentum degrees
    of freedom of each particle $i$ are displaced according to the map
    \eqref{EQriTransform} and \eqref{EQpiTransform}. The vector field
    $\eps(\rv_i)$ parameterizes the transform, which is in general
    spatially inhomogeneous and of a form such that
    $\rv_i+\eps(\rv_i)$ is a diffeomorphism.  The transformation keeps
    the differential phase space volume element $d\rvi d\pv_i$
    conserved for each particle $i$.}
\label{FIGsketchMicrostates}
\end{figure}

We assume that the shifting field $\eps(\rvi)$ has a form that
guarantees Eq.~\eqref{EQriTransform} to constitute a
diffeomorphism. This implies that: i) the transform from $\rv_i$ to
$\rv_i'$ is smooth, which follows from $\eps(\rv_i)$ being smooth, ii)
an inverse exists, such that the map from $\rv_i$ to $\rv_i'$ is
bijective, and iii) the inverse transform from $\rv_i'$ to $\rv_i$ is
also smooth.

It is straightforward to show that the Jacobian corresponding to the
transformation \eqref{EQriTransform} and \eqref{EQpiTransform} is
unity \cite{tschopp2022forceDFT, hermann2022quantum} and hence that
the particle-resolved differential phase space volume element
$d\rv_id\pv_i$ is preserved under the transformation; see
Fig.~\ref{FIGsketchMicrostates} for an illustration.  The variable
change expressed by Eqs.~\eqref{EQriTransform} and
\eqref{EQpiTransform} constitutes a canonical transformation in the
sense of classical mechanics~\cite{goldstein2002}.  This can be seen
by considering a generating function $\cal G$ from which the
transformation is obtained by differentiation
\cite{goldstein2002}. The specific form of this generator is ${\cal
  G}=\pv_i'\cdot[\rv_i + \eps(\rv_i)]$ \cite{tschopp2022forceDFT},
where the prime denotes the transformed variables. Using the generic
transformation equations $\rv_i' = \partial {\cal G}/\partial \pv_i'$
and $\pv_i = \partial {\cal G}/\partial \rv_i$ and solving for the
primed variables gives Eqs.~\eqref{EQriTransform} and
\eqref{EQpiTransform}.

Alternatively, one can verify that Eqs.~\eqref{EQriTransform} and
\eqref{EQpiTransform} constitute a canonical transformation by
explicitly computing the following Poisson bracket identities:
\begin{align}
  \{\rv'_i, \pv'_j\} &=  \delta_{ij} \unity,
  \label{EQcanonicalConditionCross}\\
  \{\rv'_i, \rv'_j\} &= \{\pv'_i, \pv'_j\} = 0,
  \label{EQcanonicalConditionSame}
\end{align}
where the Poisson bracket $\{\cdot, \cdot\}$ is expressed in the
original coordinates as $\{\hat A, \hat B\}=\sum_i [(\partial \hat
  A/\partial \rv_i)\cdot(\partial \hat B/\partial \pv_i)- (\partial
  \hat A / \partial \pv_i) \cdot (\partial \hat B / \partial \rv_i)]$,
where $\hat A(\rv^N, \pv^N)$ and $\hat B(\rv^N, \pv^N)$ are two
general phase space functions.

As a comment on the generality of the phase space shifting given by
the joint transformation \eqref{EQriTransform} and
\eqref{EQpiTransform}, this map is constructed to affect individual
particles $i$ in an identical way, which is in keeping with the aim of
describing the statistical mechanics of identical particles. The
position transform \eqref{EQriTransform} is a general smooth map
(diffeomorphism) from $\rv_i$ to $\rv_i'$. Hence the particular
additive form $\rv_i+\eps(\rv_i)$ poses no intrinsic
restrictions. Note that any map $\rv_i\to\rv_i'(\rv_i)$ can be
rewritten trivially in the form of \eqr{EQriTransform} as $\rv_i \to
\rv_i + [\rv_i'(\rv_i)-\rv_i]$, where the term in brackets then
constitutes the shifting field, $\eps(\rv_i)=\rv_i'(\rv_i)-\rv_i$.
Given the choice of position transform~\eqref{EQriTransform}, the
momentum transform \eqref{EQpiTransform} then follows uniquely from
imposing that the transformation is canonical and hence needs to
satisfy Eqs.~\eqref{EQcanonicalConditionCross} and
\eqref{EQcanonicalConditionSame}. These differential equations leave
over free remaining integration constants, which are uniquely
determined by the additional requirement that the identity
transformation is recovered in the case of vanishing shifting field,
$\eps(\rv_i) = 0$.

\subsection{Finite phase space shifting group}
\label{SECfinitePhaseSpaceShifting}

We first demonstrate that the chaining of two shifting operations
constitutes again a single shifting operation. Although this might
seem obvious based on mere geometric intuition, carrying out a direct
verification based on the structure of transformation
equations~\eqref{EQriTransform} and \eqref{EQpiTransform} is
worthwhile.  We hence consider a second shifting transform, which is
parameterized by a new shifting field $\eps_2(\rvi')$ and which acts
on the already transformed variables $\rvi'$ and $\pvi'$ that result
from applying Eqs.~\eqref{EQriTransform}
and~\eqref{EQpiTransform}. The result of the consecutive shifting is:
\begin{align}
  \rvi'' &= \rvi' + \eps_2(\rvi'),
  \label{EQrTransform2}\\
  \pvi'' &= [\unity + \nablai' \eps_2(\rvi')]^{-1}\cdot \pvi',
  \label{EQpTransform2}
\end{align}
where $\nablai'$ denotes the derivative with respect to $\rvi'$.
Replacing on the right hand sides of the transformation equations
\eqref{EQrTransform2} and \eqref{EQpTransform2} the variables $\rvi'$
and $\pvi'$ via the transformations \eqref{EQriTransform} and
\eqref{EQpiTransform} yields the following composite expressions:
\begin{align}
  \rvi'' &= \rvi + \eps_1(\rvi) + \eps_2(\rvi+\eps_1(\rvi)),
  \label{EQrTransform3}\\
  \pvi'' &= [\unity + \nablai'\eps_2(\rvi+\eps_1(\rvi))]^{-1} \cdot
  [\unity + \nablai\eps_1(\rvi)]^{-1} \cdot \pvi.
  \label{EQpTransform3}
\end{align}
We have kept the primed derivative in
$\nablai'\eps_2(\rvi+\eps_1(\rvi))$ in \eqr{EQpTransform3} to indicate
the standard gradient of the given vector field with respect to its
argument, evaluated at position $\rvi+\eps_1(\rvi)$, i.e.,
$\nablai'\eps_2(\rvi')|_{\rvi'=\rvi+\eps_1(\rvi)}$.

The conjecture that the mapping~\eqref{EQrTransform3} and
\eqref{EQpTransform3} is again a shifting transform implies that a
corresponding shifting field $\eps_{21}(\rvi)$ exists, which performs
the composite transform in a single step according to:
\begin{align}
  \rvi'' &= \rvi + \eps_{21}(\rvi),
  \label{EQrTransform4}\\
  \pvi'' &= [1+\nablai\eps_{21}(\rvi)]^{-1}\cdot\pvi.
  \label{EQpTransform4}
\end{align}
Comparing the right hand side of the composite position
transform~\eqref{EQrTransform3} with the right hand side of the
corresponding single-step position transform \eqref{EQrTransform4}
allows one to identify:
\begin{align}
  \eps_{21}(\rvi) &= \eps_1(\rvi) + \eps_2(\rvi+\eps_1(\rvi)).
  \label{EQepsPrimePrime}
\end{align}
Equation \eqref{EQepsPrimePrime} is an explicit expression for the
specific form of the shifting field $\eps_{21}(\rvi)$ that represents
the composite transform.

It thus remains to demonstrate the consistency of the single-step
momentum transform \eqref{EQpTransform4} with the
form~\eqref{EQepsPrimePrime} of the shifting field. Consistency
implies that the right hand side of Eqs.~\eqref{EQpTransform3} and
\eqref{EQpTransform4} need to be identical, which can only be true in
general provided that the following matrices are identical:
\begin{align}
  & [\unity + \nablai\eps_{21}(\rvi)]^{-1} =\notag\\&\qquad
  [\unity + \nablai'\eps_2(\rvi+\eps_1(\rvi))]^{-1}
  \cdot[\unity + \nablai\eps_1(\rvi)]^{-1}.
  \label{EQmatrixDetails1}
\end{align}
The validity of \eqr{EQmatrixDetails1} can be verified
straightforwardly by inverting the matrices on both its sides and
using the chain rule as follows. Keeping in mind to interchange the
order of the matrix product upon matrix inversion gives:
\begin{align}
  \unity + \nablai \eps_{21}(\rvi) =
  [\unity + \nablai\eps_1(\rvi)]\cdot
  [\unity + \nablai'\eps_2(\rvi+\eps_1(\rvi))].
  \label{EQmatrixDetails2}
\end{align}
Multiplying out the right hand side and simplifying, we can re-write
\eqr{EQmatrixDetails2} as $\nablai \eps_{21}(\rvi)= \nablai
\eps_1(\rvi) + \nablai'\eps_2(\rvi+\eps_1(\rvi))
+[\nablai\eps_1(\rvi)]\cdot \nablai'\eps_2(\rvi+\eps_1(\rvi))$.  This
result is identically obtained by building the gradient of
\eqr{EQepsPrimePrime} and using the chain rule on the right hand
side. This completes our proof that chaining two shifting
transformations parameterized by $\eps_1(\rv)$ and $\eps_2(\rv)$
reduces to a single shifting transformation with shifting field
$\eps_{21}(\rv)$ given by \eqr{EQepsPrimePrime}.

\begin{figure}[!t]
  \vspace{1mm}
  \includegraphics[width=.99\columnwidth]{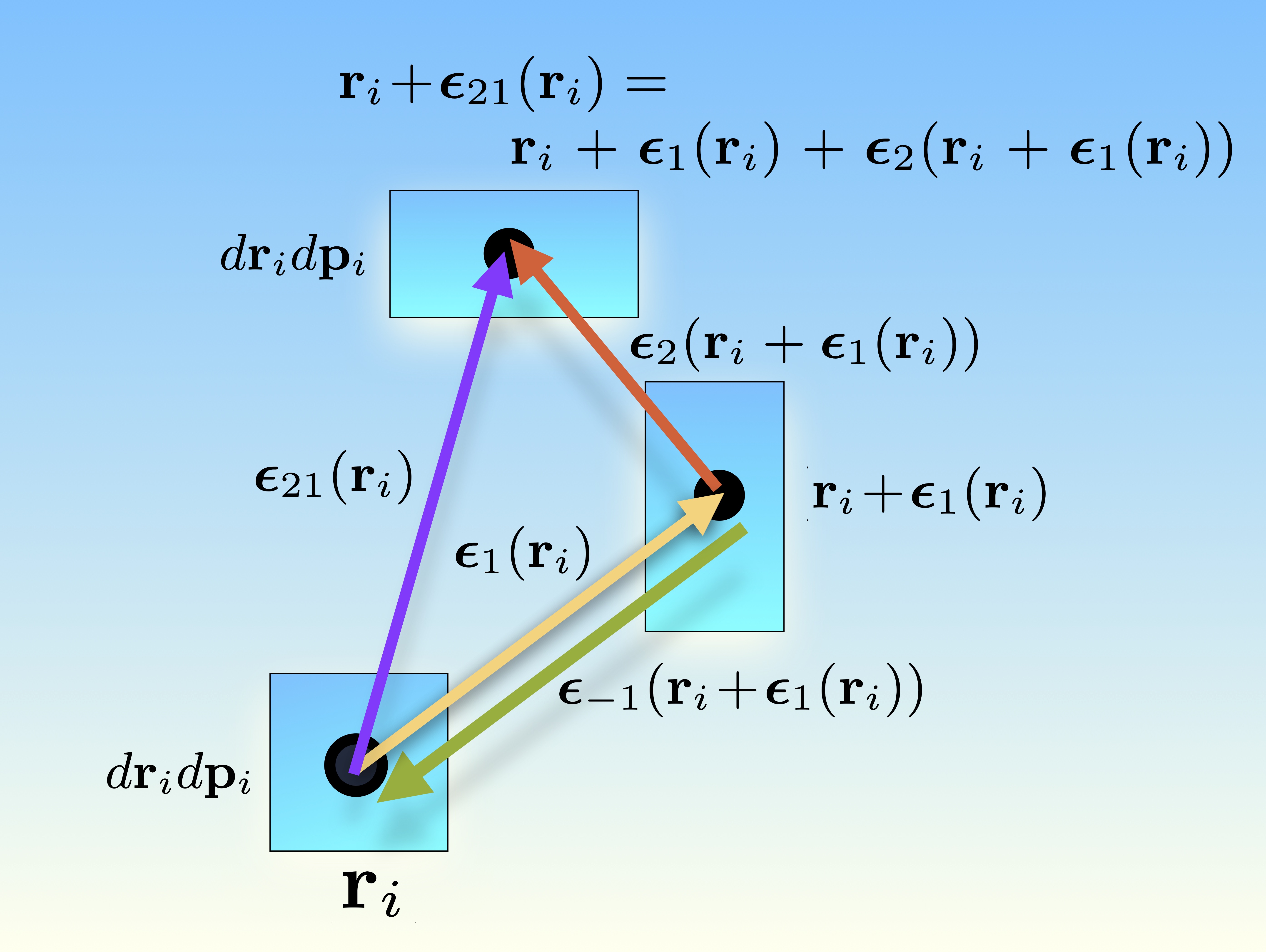}
  \caption{Group structure of local gauge transformations in
    statistical mechanics.  The rich group structure emerges when
    chaining multiple local shifting operations together. Applying two
    consecutive transformations $\eps_1(\rvi)$ and $\eps_2(\rvi)$
    results in a new transformation which is parameterized by
    $\eps_{21}(\rvi) =\eps_1(\rvi)+\eps_2(\rvi+\eps_1(\rvi))$
    according to~\eqr{EQepsPrimePrime}.  The final destination point
    is identical such that $\rvi+\eps_{21}(\rvi) = \rvi+\eps_1(\rvi)+
    \eps_2(\rvi+\eps_1(\rvi))$. For given shifting field
    $\eps_1(\rvi)$ the corresponding inverse shifting field satisfies
    $\eps_{-1}(\rvi+\eps_1(\rvi))=-\eps_1(\rvi)$ according
    to~\eqr{EQepsInverse}.  The phase space volume element $d\rvi
    d\pvi$ is conserved under the transformation, despite its changing
    shape.}
\label{FIGsketchLocal}
\end{figure}

It remains to show the existence of an inverse transform, which is
also straightforward. For given $\eps_1(\rvi)$ the inverse transform
$\eps_{-1}(\rvi)$ is obtained in the
concatenation~\eqref{EQepsPrimePrime} upon requiring that
$\eps_{-11}(\rvi)=0$, which constitutes the neutral group element
(identity) and we have replaced the index $2\to-1$.  Hence from
\eqr{EQepsPrimePrime} we obtain the inversion condition
\begin{align}
  \eps_{-1}(\rvi+\eps_1(\rvi)) = -\eps_1(\rvi),
  \label{EQepsInverse}
\end{align}
which is an implicit equation for the vector field $\eps_{-1}(\rvi)$
for given form of $\eps_1(\rvi)$. We have to restrict to vector fields
$\eps_1(\rvi)$ that allow $\eps_{-1}(\rvi)$ to exist, which is
guaranteed from the assumption that $\rvi+\eps(\rvi)$ constitutes a
diffeomorphism, as this implies the existence of an inverse. A simple
counterexample is $\eps_1(\rvi)=-\rvi$ \cite{hermann2022quantum} which
according to \eqr{EQriTransform} maps all $\rvi$ to $\rvi'=0$ and is
hence trivially not bijective.  For completeness, we exchange indices
$1$ and $-1$ in \eqr{EQepsInverse} to obtain the equivalent condition
$\eps_1(\rvi+\eps_{-1}(\rvi))=-\eps_{-1}(\rvi)$.
Figure~\ref{FIGsketchLocal} depicts an illustration of both the
chaining and the inversion of shifting.

\begin{figure}[!t]
  \vspace{1mm}
  \includegraphics[width=.99\columnwidth]{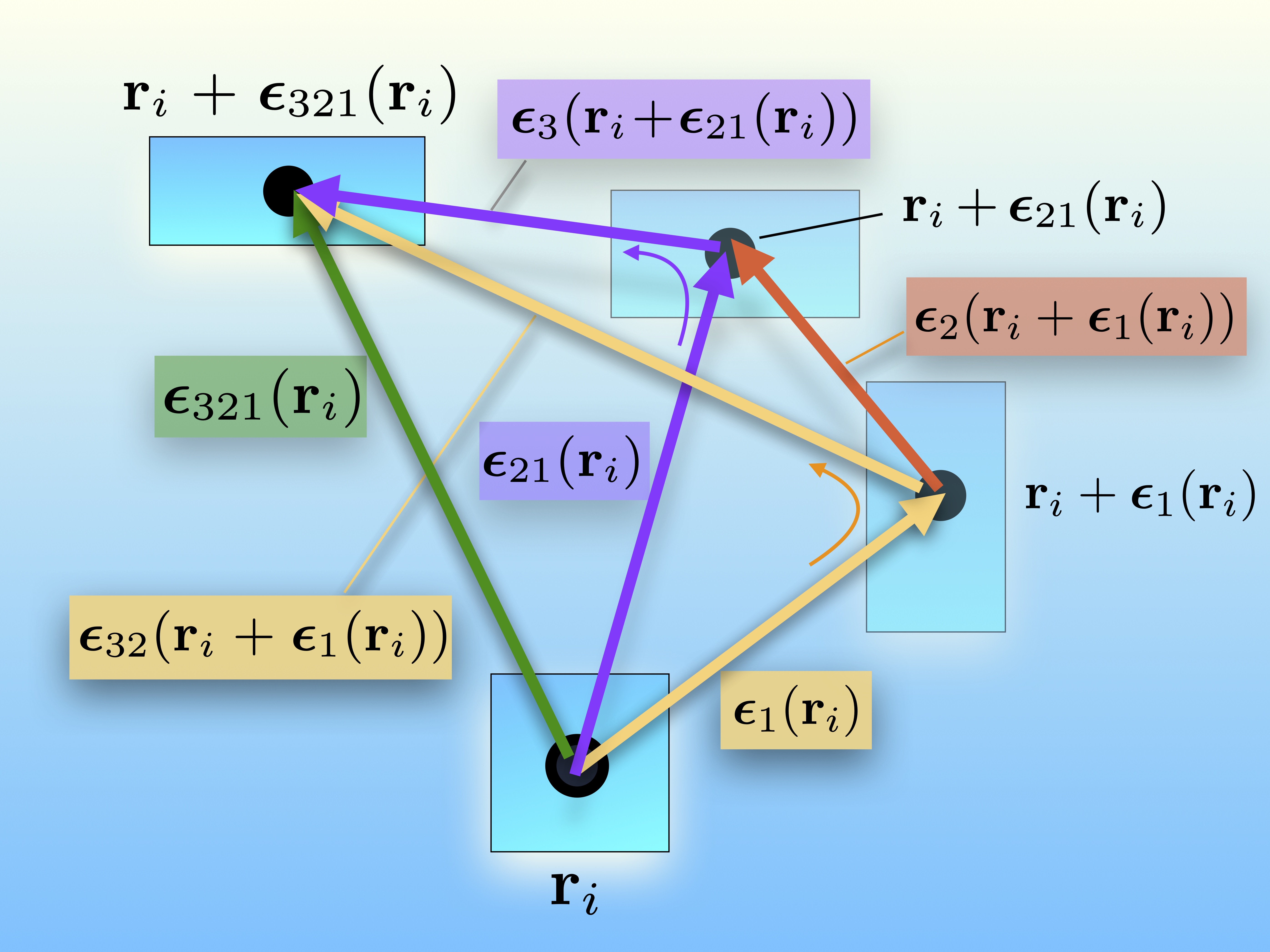}
  \caption{Associativity of the shifting transformation. Shown are
    different ways of shifting from $\rvi$ to
    $\rvi+\eps_{321}(\rv_i)$.  Verifying the associative property
    implies to show that the combined shifting vector
    $\eps_{321}(\rvi)$ of three consecutive transformations has two
    equivalent forms $\eps_{32}(\rvi+\eps_1(\rvi))+\eps_1(\rvi)=
    \eps_3(\rvi+\eps_{21}(\rvi))+\eps_{21}(\rvi)$ (indicated by the
    two curved arrows). This identity is straightforward to prove from
    applying \eqr{EQepsPrimePrime} to the two pairs of consecutive
    operations, which respectively yield
    $\eps_{21}(\rvi)=\eps_1(\rvi)+\eps_2(\rvi+\eps_1(\rvi))$ and
    $\eps_{32}(\rvi+\eps_1(\rvi))=\eps_2(\rvi+\eps_1(\rvi))
    +\eps_3(\rvi+\eps_1(\rvi)+\eps_2(\rvi+\eps_1(\rvi)))$. Inserting
    these expressions proves the associativity of shifting.}
\label{FIGsketchAssociative}
\end{figure}

Composing together more than two shifting operations is associative,
as is a requirement for valid group structure. An illustration is
shown in Fig.~\ref{FIGsketchAssociative}.  Algebraically one has to
verify that for three consecutive transforms, as are respectively
parameterized by three shifting fields $\eps_1(\rvi)$, $\eps_2(\rvi)$,
and $\eps_3(\rvi)$, the order of the grouping of pairs is irrelevant
for the final result. We recall that the non-trivial geometrical
structure arises from evaluating a subsequent shift to phase space
points that have already been displaced by the prior shift. We
demonstrate the equivalence of both groupings explicitly in the
following.  First, when applying a third transformation to the
composite~\eqref{EQepsPrimePrime}, one obtains $\rvi'''= \rvi'' +
\eps_3(\rvi'') = \rvi'' + \eps_3(\rvi+\eps_{21}(\rvi)) = \rvi' +
\eps_2(\rvi') + \eps_3(\rvi+\eps_1(\rvi) +
\eps_2(\rvi+\eps_1(\rvi)))$, which can be further made explicit as
$\rvi'''= \rvi + \eps_1(\rvi) + \eps_2(\rvi + \eps_1(\rvi)) +
\eps_3(\rvi+\eps_1(\rvi) + \eps_2(\rvi+\eps_1(\rvi)))$. On the other
hand, composing the second and third shift gives the displacement
vector $\eps_{32}(\rvi') = \eps_2(\rvi') +
\eps_3(\rvi'+\eps_2(\rvi'))$ as obtained from \eqr{EQepsPrimePrime} by
replacing $1\to 2$, $2\to 3$, and $\rvi\to\rvi'$. The overall
transformed vector is then $\rvi''' = \rvi' + \eps_{32}(\rvi') = \rvi
+ \eps_1(\rvi) + \eps_{32}(\rvi') = \rvi + \eps_1(\rvi) +
\eps_2(\rvi+\eps_1(\rvi)) +
\eps_3(\rvi+\eps_1(\rvi)+\eps_2(\rvi+\eps_1(\rvi)))$, where we have
first replaced $\rvi'$ according to \eqr{EQriTransform} and have used
the prior expression for $\eps_{32}(\rvi')$. Both expressions are
identical, which proves associativity for both the position and
momentum parts of the phase space shifting.
Figure~\ref{FIGsketchAssociative} depicts an illustration of the
associative shifting structure.

\begin{figure}[!t]
  \vspace{1mm}
  \includegraphics[width=.99\columnwidth]{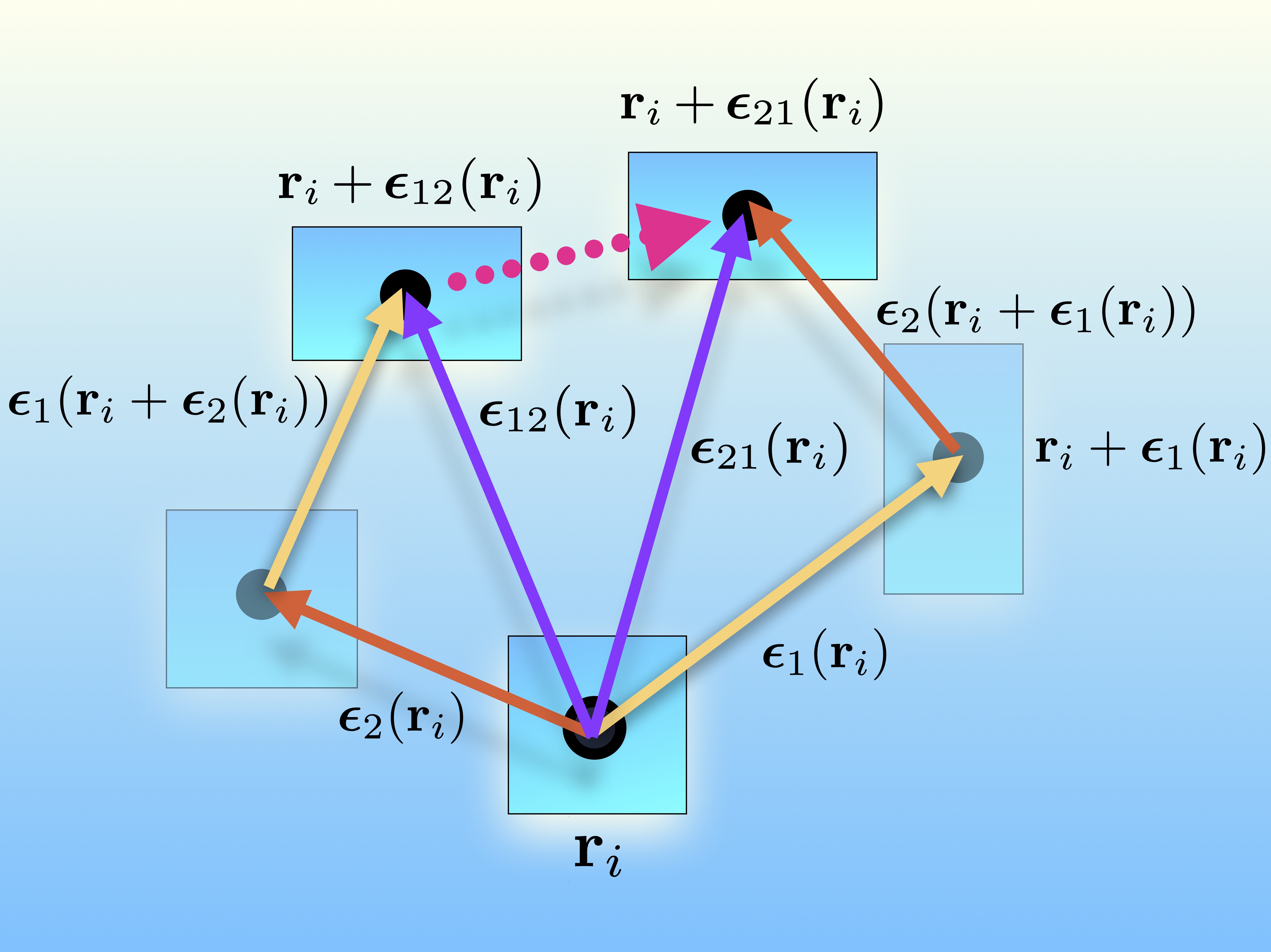}
  \caption{Non-commutative nature of the shifting transformation.
    Reversing the order of the two shifting operations that are
    respectively parameterized by $\eps_1(\rvi)$ and $\eps_2(\rvi)$
    creates a spatial mismatch (dotted arrow) between the two
    destination points $\rvi+\eps_{12}(\rvi)$ and
    $\rvi+\eps_{21}(\rvi)$.  The non-commutativity is due to the fact
    that in general $\eps_{21}(\rvi)\neq \eps_{12}(\rvi)$. Here the
    combined shifting vectors are given by
    $\eps_{21}(\rvi)=\eps_1(\rvi)+\eps_2(\rvi+\eps_1(\rvi))$,
    according to \eqr{EQepsPrimePrime}, and by
    $\eps_{12}(\rvi)=\eps_2(\rvi)+\eps_1(\rvi+\eps_2(\rvi))$, as
    obtained by exchanging the indices $1\leftrightarrow 2$.}
  \label{FIGsketchNoncommutative}
\end{figure}

Despite the fact that composition, inversion and associativity are all
relatively straightforward, the chaining of two finite shifting
operations, as specified by the composite shifting field
\eqr{EQepsPrimePrime}, is not a commutative operation. This can be
seen by interchanging the first and second shifting fields on the
right hand side of \eqr{EQepsPrimePrime}, which we reproduce for
convenience: $\eps_{21}(\rvi)= \eps_1(\rvi) +
\eps_2(\rvi+\eps_1(\rvi))$. The result is a composite shift given by
$\eps_{12}(\rvi)=\eps_2(\rvi)+ \eps_1(\rvi+\eps_2(\rvi))$, as obtained
from interchanging the indices $1\leftrightarrow 2$. In general the
two result will be different from each other, $\eps_{12}(\rv)\neq
\eps_{21}(\rv)$; see Fig.~\ref{FIGsketchNoncommutative} for an
illustration.  We demonstrate in the following
Sec.~\ref{SECinfinitesimalPositionShifting} that the non-commutative
character persists for infinitesimal shifting.

In summary, we have shown that locally resolved shifting
transformations on phase space given by Eqs.~\eqref{EQriTransform} and
\eqref{EQpiTransform} constitute a non-commutative group.  We recall
from Sec.~\ref{SECelectrodynamics} the following requirements: i) the
composite of two group element is again a group element, ii) the
existence of a neutral element, iii) the existence of an inverse, and
iv) associativity. The group elements are parameterized by the form of
their corresponding shifting vector field $\eps(\rv)$, which in
general can be of finite magnitude.

\subsection{Infinitesimal pure position shifting}
\label{SECinfinitesimalPositionShifting}

It is interesting to consider the infinitesimal version of the
shifting transformation and in particular to investigate the algebraic
structure that emerges from composite shifting.  We first address the
inversion operation which to first order in the shifting field and its
gradient is given by
\begin{align}
  \eps_{-1}(\rvi) &= 
  -\eps_1(\rvi) + \eps_1(\rvi)\cdot\nablai\eps_1(\rvi).
  \label{EQepsInverseInfinitesimal}
\end{align}
A derivation of \eqr{EQepsInverseInfinitesimal} can be based on
expanding the left hand side of the inversion condition
\eqref{EQepsInverse} to obtain
$\eps_{-1}(\rvi)+\eps_1(\rvi)\cdot\nablai\eps_{-1}(\rvi)=-\eps_1(\rvi)$,
which upon re-ordering gives $\eps_{-1}(\rvi)=-\eps_1(\rvi)
\cdot[\unity + \nablai\eps_{-1}(\rvi)]$ and
$\eps_{-1}(\rvi)\cdot[\unity+\nablai\eps_{-1}(\rvi)]^{-1}=
-\eps_1(\rvi)$. Expanding the inverse matrix and truncating after the
linear term gives $[\unity + \nablai\eps_{-1}(\rvi)]^{-1} = \unity -
\nablai\eps_{-1}(\rvi)$ which yields \eqr{EQepsInverseInfinitesimal}
upon interchanging $\eps_{-1}(\rvi)$ and $\eps_1(\rvi)$.

We consider the action of the spatial displacement
\eqref{EQriTransform} on a position-dependent function $f(\rvi)$,
which is hence moved to $f(\rvi')=f(\rvi+\eps_1(\rvi))$, as obtained
from using the position transform \eqref{EQriTransform}. Taylor
expanding in $\eps_1(\rvi)$ yields $f(\rvi')=
f(\rvi)+\eps_1(\rvi)\cdot\nablai f(\rvi)$ to linear order in
$\eps_1(\rvi)$. From the structure of the Taylor expansion we can
identify the differential operator $\eps_1(\rvi)\cdot\nablai$ as
performing the change of the function $f(\rvi)$ to $f(\rvi')$ to first
order in $\eps_1(\rvi)$. Correspondingly, shifting by a vector field
$\eps_2(\rvi)$ is represented by the operator
$\eps_2(\rvi)\cdot\nablai$.

Successively applying two shifts requires to first expand $f(\rvi'') =
f(\rvi'+\eps_2(\rvi'))$ around $\rvi'$ and then, after inserting
$\rvi' = \rvi + \eps_1(\rvi)$, to expand around $\rvi$.  The overall
result is $f(\rvi'')= [f(\rvi) + \eps_2(\rvi)\cdot\nablai f(\rvi)] +
\eps_1(\rvi)\cdot\nablai [f(\rvi) + \eps_2(\rvi)\cdot\nablai
  f(\rvi)]$, which we re-group as:
\begin{align}
  f(\rvi'') &= 
  f(\rvi)+[\eps_1(\rvi)+\eps_2(\rvi)]\cdot\nablai f(\rvi)  \notag\\&\quad
  + \eps_1(\rvi)\cdot \nablai [\eps_2(\rvi)\cdot \nablai f(\rvi)].
  \label{EQsuccessivePositionShift}
\end{align}
Here the order of the two consecutive shifts is imprinted in the
specific form of the last term of the sum on the right hand side of
\eqr{EQsuccessivePositionShift}. Reversing the order of the two shifts
leads to a corresponding term $\eps_2(\rvi)\cdot\nablai
[\eps_1(\rvi)\cdot\nablai f(\rvi)]$, which in general is different
from the last term in \eqr{EQsuccessivePositionShift}.

To systematically capture the emerging difference it is appropriate to
consider the commutator of two shifting differential operators
$\eps_1(\rvi)\cdot\nablai$ and $\eps_2(\rvi)\cdot\nablai$.  The
following structure is obtained from straightforward calculation:
\begin{align}
  [\eps_1(\rvi)\cdot\nablai, \eps_2(\rvi)\cdot\nablai] &=
  \eps_\Delta(\rvi)\cdot\nablai,
  \label{EQshiftingCommutator}
\end{align}
where the brackets denote the commutator of two operators ${\cal O}_1$
and ${\cal O}_2$ in the usual way as $[{\cal O}_1, {\cal O}_2] = {\cal
  O}_1 {\cal O}_2- {\cal O}_2{\cal O}_1$.  Hence spelling out
\eqr{EQshiftingCommutator} explicitly gives
$[\eps_1(\rvi)\cdot\nablai, \eps_2(\rvi)\cdot\nablai] =
\eps_1(\rvi)\cdot\nablai \eps_2(\rvi)\cdot\nablai -
\eps_2(\rvi)\cdot\nablai \eps_1(\rvi)\cdot\nablai$, with each
derivative $\nablai$ acting on all functions to its right.  

The vector field $\eps_\Delta(\rv)$ in \eqr{EQshiftingCommutator} that
represents the difference between the two composite shifts is given by
\begin{align}
  \eps_\Delta(\rvi) &=
  \eps_1(\rvi)\cdot[\nablai\eps_2(\rvi)]
  -\eps_2(\rvi)\cdot[\nablai\eps_1(\rvi)]
  \label{EQepsDelta}\\
  &=  [\eps_1(\rvi), \eps_2(\rvi)]_L,
  \label{EQLieBracketDefinition}
\end{align}
where the matrices $[\nablai\eps_1(\rvi)]$ and $[\nablai\eps_2(\rvi)]$
in \eqr{EQepsDelta} are not differential operators but rather tensor
fields. In general the right hand side of \eqr{EQepsDelta} will not
vanish, $\eps_\Delta(\rvi)\neq 0$. The particular combination
\eqref{EQepsDelta} of two given vector fields $\eps_1(\rvi)$ and
$\eps_2(\rvi)$ constitutes the {\it Lie bracket} $[\eps_1(\rvi),
  \eps_2(\rvi)]_L$ as is defined in the standard way
\cite{robbin2022book} in \eqr{EQepsDelta}.

Using the Lie bracket \eqref{EQLieBracketDefinition} we can rewrite
the right hand side of the shifting commutator relation
\eqref{EQshiftingCommutator} to express the commutator relationship
as:
\begin{align}
  [\eps_1(\rvi)\cdot\nablai, \eps_2(\rvi)\cdot\nablai] &=
  [\eps_1(\rvi), \eps_2(\rvi)]_L\cdot\nablai,
  \label{EQshiftingCommutatorViaLieBracket}
\end{align}
which constitutes a Lie algebra of differential operators.

We have so far formulated the theory on the basis of specific choices
of the shifting fields $\eps_1(\rv)$, $\eps_2(\rv)$ and the resulting
form of $\eps_\Delta(\rv)$. However, the specific form of these vector
fields is not of physical interest, as any observables are independent
of the shifting fields, as is consistent with their role as mere gauge
functions.

Separating the specificities from the intrinsic nature of the shifting
is technically achieved by working with functional differentiation
(see Ref.~\cite{schmidt2022rmp} for a practitioners' account) and we
recall that Noether's theorem addresses such invariant variations
\cite{noether1918}. Hence having identified the position shifting
operators $\eps(\rvi)\cdot\nabla_i$, we functionally differentiate
with respect to $\eps(\rv)$ as a measure of the response of the
shifting operator $\eps(\rvi)\cdot\nabla_i$ against changes in the
shifting field at a (new) position $\rv$.

The functional derivative is straightforward to carry out and leads
to
\begin{align}
  \frac{\delta}{\delta\eps(\sv)} \eps(\rvi)\cdot\nablai
  &= \delta(\rvi-\sv) \nablai,
  \label{EQfuctionalDerivativeOfEpsNabla}
\end{align}
where we have used that $\delta\eps(\rvi)/\delta\eps(\sv)=\unity
\delta(\rvi-\sv)$ and $\unity\cdot\nabla=\nabla$.  Multiplying (via a
scalar product) the right hand side of
\eqr{EQfuctionalDerivativeOfEpsNabla} with $\eps(\sv)$ and integrating
over $\sv$ recovers the original position-shifting operator according
to:
\begin{align}
  \eps(\rvi)\cdot\nablai &= 
  \int d\sv \eps(\sv) \cdot \delta(\rvi-\sv) \nabla_i.
  \label{EQepsNablaFromIntegration}
\end{align}
Equation \eqref{EQepsNablaFromIntegration} can be viewed as a simple
case of functional integration (where here the functional integrand is
independent of the functional argument $\eps(\rv)$).  We refer to
Refs.~\cite{evans1979, evans1992, sammueller2023whyNeural,
  sammueller2024hyperDFT, sammueller2024whyhyperDFT,
  sammueller2024attraction} both for descriptions and use of
functional integration in more general settings, in particular in the
context neural functionals \cite{sammueller2023whyNeural,
  sammueller2024hyperDFT, sammueller2024whyhyperDFT,
  sammueller2024attraction}.

The commutator relationship \eqref{EQshiftingCommutatorViaLieBracket}
has a simple bilinear dependence on the two shifting fields. It is
hence straightforward to build the mixed second functional derivative
of \eqr{EQshiftingCommutator} according to
$\delta^2/[\delta\eps_1(\sv)\delta\eps_2(\sv')]$, where $\sv$ and
$\sv'$ are two new position variables. The functional dependence on
the shifting fields is explicit on the left hand side of
\eqr{EQshiftingCommutatorViaLieBracket}. Functionally differentiating
the right hand side of \eqr{EQshiftingCommutatorViaLieBracket}
requires to take into account that $\eps_\Delta(\rvi)$ depends on the
vector fields $\eps_1(\rvi)$ and $\eps_2(\rvi)$ via \eqr{EQepsDelta},
which gives $\delta^2 \epsilon_{\Delta c}(\rvi) /
[\delta\epsilon_{1a}(\sv) \delta\epsilon_{2b}(\sv') ] =
\delta(\rvi-\sv) [\nablai^a\delta(\rvi-\sv')\delta_{bc}]
-\delta(\rvi-\sv') [\nablai^b \delta(\rvi-\sv)\delta_{ac}] $, where
$a,b,c$ denote the Cartesian components and we have used $\delta
\epsilon_a(\rvi)/\delta\epsilon_b(\sv)= \delta(\rvi-\sv)\delta_{ab}$.
The overall result for the mixed functional derivative of the
commutator relationship \eqref{EQshiftingCommutatorViaLieBracket} is:
\begin{align}
  &[\delta(\rvi-\sv)\nablai, \delta(\rvi-\sv')\nablai]
  \notag\\ & \;
  =[\nablai\delta(\rvi-\sv')] \delta(\rvi-\sv) \nablai
  -[\nablai \delta(\rvi-\sv)] \delta(\rvi-\sv') \nablai^{\sf T}\\
  &\;
  = \delta(\rvi-\sv')\nablai [\nabla_{}\delta(\sv-\sv')]
  + [\nabla_{}\delta(\sv-\sv')] \delta(\rvi-\sv)\nablai,
  \label{EQprotoVersionsigmacCommutator}
\end{align}
where as before $\nabla_{}$ indicates the derivative with respect to
$\rv$ and the superscript $\sf T$ denotes matrix transposition, here
of a dyadic product.

We introduce an integral over $\sv''$ on the right hand to express the
result as the following operator identity
\begin{align}
  &[\delta(\rvi-\sv)\nablai, \delta(\rvi-\sv')\nablai]
  \notag\\ & \qquad
  =\int d\sv'' {\sf D}(\sv,\sv',\sv'') 
  \cdot \delta(\rvi-\sv'')\nablai.
  \label{EQlieAlgebraQ}
\end{align}
where ${\sf D}$ is a third-rank tensor that plays the role of Lie
structure constants. The $abc$-component of ${\sf D}$ is given by
\begin{align}
  D_{abc}(\sv,\sv',\sv'') &= 
  \delta_{bc} \delta(\sv''-\sv)[\nabla''_{a}\delta(\sv''-\sv')] 
  \notag\\&\quad
  -\delta_{ac}\delta(\sv''-\sv')[\nabla''_{b}\delta(\sv''-\sv)],
  \label{EQstructureConstantsPositionShifting}
\end{align}
where $\nabla_{}''$ denotes the derivative with respect to $\sv''$.

In summary we have demonstrated, for the case of position-only
shifting, the relationship of the spatial displacement with the action
of shifting differential operators. We have shown that the
non-commutative nature that is inherent when considering finite
displacements persists when Taylor expanding and keeping only the
lowest nontrivial terms. The non-commutative aspect is then reflected
by algebraic relationships. Specifically, when expressed in the form
\eqref{EQstructureConstantsPositionShifting} it is apparent that the
particular form of the product of the delta distribution and
derivative operator, $\delta(\rvi-\sv)\nablai$, persists as a result
of building the commutator.

In Sec.~\ref{SECconfigurationShifting}, we demonstrate that much of
this favourable structure can be retained when considering the joint
phase space transformation of coordinates and momenta. We recall that
this constitutes a canonical transformation, which the present
position-only shifting with unchanged momenta does not.

\section{Statistical Mechanics}
\label{SECstatisticalMechanics}

\subsection{Partition sum and thermal averages}
\label{SECpartitionSumAndThermalAverages}

We present the invariance argument \cite{robitschko2024any} for the
thermal mean of a phase space function $\hat A(\rv^N)$, where
$\rv^N=\rv_1,\ldots, \rv_N$ is a shorthand for all position
coordinates in the system with $\rv_i$ denoting the position
coordinate of particle $i=1,\ldots,N$ in $d$ spatial dimensions, and
$N$ indicates the total number of particles. The canonical average is
given by the classical trace operation $\Tr \cdot =
(N!h^{dN})^{-1}\int d\rv^N d\pv^N \cdot$, where $h$ indicates Planck's
constant and $\int d\rv^N d\pv^N$ denotes the phase space integral
over all coordinates $\rv^N$ and momentum variables $\pv^N=\pv_1,
\ldots, \pv_N$. The partition sum is $Z=\Tr \e^{-\beta
  H(\rv^N,\pv^N)}$, where $H(\rv^N,\pv^N)$ denotes the Hamiltonian,
with inverse temperature $\beta=1/(k_BT)$, where $k_B$ denotes the
Boltzmann constant and $T$ the absolute temperature. The Helmholtz
free energy is then $F=-k_BT\ln Z$. Thermal averages are obtained as 
\begin{align}
  \langle \cdot \rangle = \Tr
  \cdot \e^{-\beta H(\rv^N,\pv^N)}/Z.
  \label{EQaverageDefinition}
\end{align}

More explicitly the thermal average of an observable $\hat A(\rv^N)$
is given by
\begin{align}
   A& =
  \frac{1}{h^{dN}N!}\int d\rv^N d\pv^N \hat A(\rv^N) 
  \e^{-\beta H(\rv^N, \pv^N)}/Z
  \label{EQmeanA1}
\end{align}
We first rewrite the phase space integral in \eqr{EQmeanA1} in an
identical way upon merely renaming the unprimed by primed variables:
\begin{align}
  A &= \frac{1}{h^{dN}N!} \int d\rv'^N d\pv'^N 
  \hat A(\rv'^N)
  \e^{-\beta H(\rv'^N, \pv'^N)}/Z',
  \label{EQmeanA2}
\end{align}
where the primed partition sum $Z'$ is expressed in the primed
variables and $Z'=Z$.

We next perform the canonical variable transformation given by
Eqs.~\eqref{EQriTransform} and \eqref{EQpiTransform}. The phase space
volume element is thereby conserved such that $d \rv_i
d\pv_i=d\rv_i'd\pv_i'$ for each particle $i$. The result is
\begin{align}
  A  &= \frac{1}{h^{dN}N!} \int d\rv^N d\pv^N 
  \hat A(\rv'^N)
  \e^{-\beta H(\rv'^N,\pv'^N)} / Z',
  \label{EQmeanA3}  \\
  &= A[\eps],
  \label{EQmeanA4}
\end{align}
The dependence of the integrand in \eqr{EQmeanA3} on the unprimed
variables is suppressed in the notation for clarity and it occurs via
the transformation equations \eqref{EQriTransform} and
\eqref{EQpiTransform} that render the $\rv_i'$ and $\pv_i'$
parametrically dependent on $\rv_i$ and $\pv_i$ and functionally
dependent on the form of the shifting field $\eps(\rv)$.  Spelling out
these dependencies explicitly we have $\rv_i'=\rv_i'(\rv_i,[\eps])$
according to \eqr{EQriTransform} and $\pv_i'=\pv'(\rv_i,\pv_i,[\eps])$
according to \eqr{EQpiTransform}.

After the unprimed phase space integral is performed, at face value
\eqr{EQmeanA3} depends functionally on the shifting field, which we
have indicated as $A[\eps]$ in \eqr{EQmeanA4}. Comparing
\eqr{EQmeanA1} with \eqr{EQmeanA4} gives the invariance condition
$A=A[\eps]$, where the left hand side can also be understood as
$A[\eps]$ being evaluated at vanishing shifting field, $\eps(\rv)=0$,
such that $A[0]=A[\eps]$.

As the phase space shifting affects systems with fixed number of
particles, the argumentation carries over to the grand ensemble where
each system with $N$ particles is displaced identically.  Hence
besides the changes in the definition of the grand ensemble averages,
no further alterations occur in any of the results.

To be specific, we consider Hamiltonians that consist of kinetic,
interparticle, and external energy contributions according to the sum:
\begin{align}
  H(\rv^N, \pv^N) &= \sum_i\frac{\pv_i^2}{2m} 
  + u(\rv^N) + \sum_i V_\rmext(\rv_i),
  \label{EQHamiltonian}
\end{align}
where the sums run over all $N$ particle indices $i$, $m$ denotes the
particle mass, $u(\rv^N)$ is the interparticle interaction potential,
and $V_\rmext(\rv)$ is an external one-body potential.

\subsection{One-body observables}
\label{SEConeBodyObservables}

We lay out the relevant one-body observables, which are pertinent in
the equilibrium sum rules that follow from the gauge invariance as
described above. The one-body level of correlation functions is
integral in the classical density functional formulation of
statistical mechanics \cite{evans1979, evans1992, evans2016,
  hansen2013, schmidt2022rmp} and it naturally extends to the
dynamical problems, see Ref.~\cite{schmidt2022rmp} for systematic
(power functional) derivations on the basis of classical Hamiltonian,
overdamped Brownian, and quantum many-body time evolution.

The one-body localization enables one to carry out systematic
coarse-graining in the sense of reduction of the full phase space
information, while it remains microscopically sharp, such that all
features on microscopic length scales are resolved without any
principal loss. The primary one-body observable is the one-body
density distribution or density ``profile'' $\rho(\rv)$. When working
with functional relationships two- and higher-body observables are in
principle accessible via functional differentiation both in
\cite{evans1979, evans1992, evans2016, hansen2013, schmidt2022rmp} and
out-of-equilibrium \cite{schmidt2022rmp}.

As an aside, to reach beyond the hierarchy of increasingly
higher-order correlation functions, the recent hyperdensity functional
theory \cite{sammueller2024hyperDFT, sammueller2024whyhyperDFT} allows
one to treat complex order parameters, including algorithmically
defined phase space functions, on the basis of the density profile as
the fundamental variable, as is ensured by the Mermin-Evans
\cite{mermin1965, evans1979} density functional map.

The response of the Hamiltonian $H$ against changes in the form of the
external potential $V_\rmext(\rv)$ is measured by the functional
derivative $\hat\rho(\rv)=\delta H / \delta V_\rmext(\rv)$, where the
density ``operator'' (phase space function) is obtained as:
\begin{align}
  \hat\rho(\rv) &= \sum_i \delta(\rv-\rv_i).
  \label{EQdensityOperator}
\end{align}
The density profile is the thermal average, $\rho(\rv)=\langle
\hat\rho(\rv)\rangle$. Its use allows one to express the average
localized external force density as $-\rho(\rv)\nabla V_\rmext(\rv)$,
where we recall that $V_\rmext(\rv)$ is the external potential that a
particle situated at position $\rv$ contributes, and $-\nabla
V_\rmext(\rv)$ is the corresponding external force field. It is
straightforward to show the equality $-\rho(\rv)\nabla
V_\rmext(\rv)=\langle \sum_i
\delta(\rv-\rv_i)\fv_\rmext(\rv_i)\rangle$, where the external force
field is $\fv_\rmext(\rv)=-\nabla V_\rmext(\rv)$.

The localized interparticle force density ``operator'' (phase space
function) is defined as:
\begin{align}
  \hat\Fv_\rmint(\rv) &=
  -\sum_i \delta(\rv-\rv_i) \nabla_i u(\rv^N).
  \label{EQFintOperatorDefinition}
\end{align}
The corresponding mean interparticle force density is simply the
average $\Fv_\rmint(\rv)=\langle \hat \Fv_\rmint(\rv) \rangle$, with
$\hat \Fv_\rmint(\rv)$ defined by
\eqr{EQFintOperatorDefinition}. Interparticle and external force
densities can be combined into a dedicated potential force density:
\begin{align}
  \Fv_U(\rv) &= \Fv_\rmint(\rv) - \rho(\rv) \nabla V_\rmext(\rv),
  \label{EQFUDefinition}
\end{align}
where the subscript $U$ refers to total potential energy, as
represented by the corresponding phase space function $\hat
U=u(\rv^N)+\sum_iV_\rmext(\rv_i)$.  The resulting potential force
density ``operator'' is analogously defined via
\begin{align}
  \hat\Fv_U(\rv) &= \hat\Fv_\rmint(\rv) -
  \hat\rho(\rv) \nabla V_\rmext(\rv),
  \label{EQFUOperator}
\end{align}
and by construction the average is the mean potential force density,
$\Fv_U(\rv)=\langle \hat\Fv_U(\rv) \rangle$.

In equilibrium the potential forces are balanced by a thermal force
density which arises from the kinetic energy contribution to the
Hamiltonian \eqref{EQHamiltonian}. The general phase space function
that underlies this effect is the localized kinetic stress
``operator'' $\hat\taub(\rv)$ which is defined as:
\begin{align}
  \hat\taub(\rv) &= 
  -\sum_i \frac{\pv_i\pv_i}{m} \delta(\rv-\rv_i),
  \label{EQkineticStressOperator}
\end{align}
where $\pv_i\pv_i$ denotes the dyadic product of the momentum of
particle $i$ with itself. It is straightforward to show upon using the
properties of the Maxwellian, i.e., the Gaussian distribution of the
momenta according to the Boltzmann factor, that
$\nabla\cdot\taub(\rv)=\nabla\cdot \langle \hat\taub(\rv) \rangle =
-k_BT \nabla\rho(\rv)$, which acts as a thermal, diffusive force
density.

Collecting kinetic, interparticle, and external contributions, we can
define a total force density operator:
\begin{align}
  \hat\Fv(\rv) &= 
  \nabla\cdot\hat\taub(\rv)
  +\hat\Fv_\rmint(\rv)
  -\hat\rho(\rv)\nabla V_\rmext(\rv),
  \label{EQforceDensityOperator}
\end{align}
where we refer the Reader to Ref.~\cite{schmidt2022rmp} for the
derivation from the second time derivative of the density operator
\eqref{EQdensityOperator}.  In thermal equilibrium the average
one-body force density $\Fv(\rv)=\langle \hat\Fv(\rv) \rangle$
vanishes,
\begin{align}
  \Fv(\rv) &= 0,
  \label{EQforceDensityBalanceEquilibrium}
\end{align}
which upon spelling out the three individual terms is analogous to the
more detailed form:
\begin{align}
  -k_BT \nabla\rho(\rv) + \Fv_\rmint(\rv)
  -\rho(\rv) \nabla V_\rmext(\rv)
  &= 0.
\end{align}
The equilibrium force density balance
\eqref{EQforceDensityBalanceEquilibrium} can be verified by explicit
calculation of the thermal averages upon using partial integration on
phase space to manipulate the occurring coordinate derivatives.  In
particular when the interparticle force density $\Fv_\rmint(\rv)$ is
expressed as an integral over the two-body density multiplied by the
pair force, \eqr{EQforceDensityBalanceEquilibrium} is commonly
referred to as the Yvon-Born-Green (YBG) equation \cite{yvon1935,
  born1946, hansen2013}.

Finally, it is useful to define a configurational force density
``operator'' as
\begin{align}
  \beta \hat\Fv_c(\rv) &= -\nabla \hat \rho(\rv) + \beta \hat\Fv_U(\rv),
  \label{EQconfigurationalForceDensityOperator} 
\end{align}
which on average $\langle\hat\Fv_c(\rv)\rangle = 0$, which is
identical to the equilibrium force density
balance~\eqref{EQforceDensityBalanceEquilibrium}. Note that for purely
configurational observables $\hat A(\rv^N)$, due to the simplicity of
the Maxwellian, we have $\langle \hat A(\rv^N)\hat\Fv_c(\rv)\rangle =
\langle \hat A(\rv^N) \hat \Fv(\rv)\rangle$, see
e.g.\ Ref.~\cite{hermann2023whatIsLiquid} for further details of the
thermal average of the kinetic stress operator
\eqref{EQkineticStressOperator}.

\subsection{Sum rules from pure position shifting}
\label{SECconfigurationShifting}

The mathematical structure of infinitesimal position shifting
described in Sec.~\ref{SECinfinitesimalPositionShifting} can be
directly used to derive statistical mechanical sum rules, as we
demonstrate in the following. We define the following localized
position-shifting operators
\begin{align}
  \bsig_c(\rv) = \sum_i \delta(\rv-\rv_i) \nabla_i,
  \label{EQsigmacDefinition}
\end{align}
where the subscript $c$ indicates that only the configuration $\rv^N$
is affected, independent of the momentum degrees of freedom $\pv^N$.

The adjoint operator is given by
\begin{align}
  \bsig_c^\dagger(\rv) &= -\bsig_c(\rv) + \nabla\hat\rho(\rv),
  \label{EQsigmacAdjoint}
\end{align}
where the density operator $\hat\rho(\rv)$ is defined via
\eqr{EQdensityOperator}. Equation \eqref{EQsigmacAdjoint} follows from
configuration space (position) integration by parts, the product rule,
and noting that in the resulting expression one can replace
$-\nabla_i=\nabla$. Here the adjoint is understood with respect to the
phase space integral, such that for any two general phase space
functions $f(\rv^N, \pv^N)$ and $g(\rv^N, \pv^N)$ the adjoint ${\cal
  O}^\dagger$ of a given operator ${\cal O}$ is defined as satisfying
$\int d\rv^N d\pv^N f {\cal O}^\dagger g = \int d\rv^N d\pv^N g {\cal
  O} f$.

The configurational shifting operators at two generic positions~$\rv$
and $\rv'$ satisfy the commutator relationship
\begin{align}
    [\bsig_c(\rv), \bsig_c(\rv')] &=
    \bsig_c(\rv')[\nabla\delta(\rv-\rv')]
    +[\nabla\delta(\rv-\rv')] \bsig_c(\rv),
    \label{EQsigmacCommutator}
\end{align}
Equation~\eqref{EQsigmacCommutator} follows straightforwardly from the
corresponding single-particle commutator
\eqref{EQprotoVersionsigmacCommutator} upon summing the latter over
$i$, then identifying the explicit form~\eqref{EQsigmacDefinition}
of~$\bsig_c(\rv)$, and noticing that the commutator vanishes for two
different particles $i\neq j$. We use the prime in the two different
roles to denote the generic position variable~$\rv'$, which is in
general unrelated to $\rv$, as well as to indicate the transformed
phase space coordinate $\rv_i'$, which is related to $\rvi$ via the
position transform \eqref{EQriTransform}.

It is interesting to apply the localized configuration shift operator
\eqref{EQsigmacDefinition} to the Hamiltonian, which yields
\begin{align}
  -[\bsig_c(\rv) H] &= \hat\Fv_U(\rv),
  \label{EQFUOperatorFromShifting}
\end{align}
where we recall the potential force density ``operator''
$\hat\Fv_U(\rv)$ as the sum \eqref{EQFUOperator} of interparticle and
external contributions. The brackets on the left hand side of
\eqr{EQFUOperatorFromShifting} indicate the range of action of
$\bsig_c(\rv)$, i.e., it only operates on $H$ but not beyond. The
explicit verification of \eqr{EQFUOperatorFromShifting} can be based
on the definition \eqref{EQsigmacDefinition} of $\bsig_c(\rv)$, the
form \eqref{EQHamiltonian} of the Hamiltonian, and the definition
\eqref{EQFUDefinition} of the potential force operator
$\hat\Fv_U(\rv)$.

Similarly, when applied to the Boltzmann factor, the chain rule
together with \eqr{EQFUOperatorFromShifting} yields straightforwardly
\begin{align}
  [\bsig_c(\rv) \e^{-\beta H}]&= \beta \hat\Fv_U(\rv) \e^{-\beta H}.
  \label{EQFUOperatorFromBoltzmannShifting}
\end{align}
Equation \eqref{EQFUOperatorFromBoltzmannShifting} constitutes a
fundamental link between spatial shifting of the (equilibrium)
many-body probability distribution and the emergence of the localized
force density operator. It remains to exploit the specific operator
structure of the localized shifting operators, as embodied in
Eqs.~\eqref{EQsigmacAdjoint} and \eqref{EQsigmacCommutator}, and to
build the average over the equilibrium ensemble.

To derive hyperforce sum rules we multiply the adjoint identity
\eqref{EQsigmacAdjoint} from the left by a purely coordinate-dependent
phase space function $\hat A(\rv^N)$ and then build the thermal
average of the result, which gives:
\begin{align}
  \langle \hat A \bsig_c^\dagger(\rv) \rangle
  &= -\langle \hat A \bsig_c(\rv) \rangle
  + \nabla \langle\hat A \hat\rho(\rv)\rangle.
  \label{EQsigmacAdjointAverage}
\end{align}
Recalling the thermal average being defined via
\eqr{EQaverageDefinition} the left hand side of
\eqr{EQsigmacAdjointAverage} is explicitly given by $ \langle \hat A
\bsig_c^\dagger(\rv) \rangle = \Tr \hat A \bsig_c^\dagger(\rv )
\e^{-\beta H}/Z= \Tr [\bsig_c(\rv)\hat A]\e^{-\beta H}/Z = \langle
  [\bsig_c(\rv) \hat A]\rangle$, where in the second step we have made
  use of the defining property of the adjoint operator.
Making also the first term on the right hand side of
\eqr{EQsigmacAdjointAverage} more explicit yields $-\langle \hat A
\bsig_c(\rv) \rangle = -\Tr \hat A \bsig_c(\rv) \e^{-\beta H}/Z = -\Tr
\hat A \beta \hat\Fv_U(\rv) \e^{-\beta H}/Z= -\langle \hat A \beta
\hat\Fv_U(\rv)\rangle$, where we have first used the chain rule and
then applied the configuration shift operator to the Boltzmann factor
according to \eqr{EQFUOperatorFromBoltzmannShifting}.

As the second term on the right hand side of
\eqr{EQsigmacAdjointAverage} is already in the form of an explicit
average, we can collect all terms to formulate the following exact
equilibrium sum rule
\begin{align}
  \Sv_{A}(\rv)  &= 
  -\langle \hat A \beta \hat\Fv_U(\rv)\rangle 
  + \nabla \langle \hat A \hat\rho(\rv)\rangle,
  \label{EQsigmacAdjointAverage2} 
\end{align}
where the hyperforce density $\Sv_A(\rv) = \langle
\hat\Sv_A(\rv)\rangle$ is given in the present case of mere dependence
on coordinates (independent of momenta) as
\begin{align}
  \hat \Sv_A(\rv) &= [\bsig_c(\rv) \hat A(\rv^N)]
  \label{EQSAconfigDefinition}
  \\  
  &= \sum_i \delta(\rv-\rv_i)
  [\nabla_i \hat A(\rv^N)].
  \label{EQsigmacAdjointAverage3} 
\end{align}
The form \eqref{EQsigmacAdjointAverage3} follows directly from the
definition of the position shift operator \eqref{EQsigmacDefinition}.
Using the configurational force density operator $\hat\Fv_c(\rv)$
given by \eqr{EQconfigurationalForceDensityOperator} and rearranging
allows one to rewrite the sum rule \eqr{EQsigmacAdjointAverage2} as
\begin{align}
  \Sv_A(\rv) + \langle \hat A(\rv^N) \beta\hat\Fv_c(\rv) \rangle &= 0.
  \label{EQsigmacAdjointAverage4}
\end{align}

The situation becomes even richer when using two (or more) shifting
operators due to a larger variety of re-writing of equivalent
expressions. This multitude could potentially be useful in concrete
applications. We here restrict ourselves to two specific cases and
hence first consider the operator algebra
\eqref{EQsigmacCommutator}. To arrive at a sum rule, we multiply
\eqr{EQsigmacCommutator} by $\hat A(\rv^N)$ from the left and then
build the thermal average. The first term of the commutator is
$\langle \hat A \bsig_c(\rv)\bsig_c(\rv')\rangle = \langle
[\bsig_c^\dagger(\rv)\hat A]\bsig_c(\rv')\rangle =
\langle[-\bsig_c(\rv)\hat A + \nabla \hat\rho(\rv)\hat A]\beta
\hat\Fv_U(\rv')\rangle=-\langle\hat \Sv_A(\rv)\beta\hat\Fv_U(\rv')
\rangle+\nabla \langle \hat A \hat\rho(\rv) \beta
\hat\Fv_U(\rv')\rangle$, where we have first re-written via the
adjoint $\bsig_c^\dagger(\rv)$, and then used its relationship
\eqref{EQsigmacAdjoint} and the definition
\eqref{EQSAconfigDefinition} of the configurational hyperforce density
operator $\Sv_A(\rv)$.

The result for the second term of the commutator either follows from
an analogous chain of steps or, equivalently, via simply exchanging
$\rv$ and $\rv'$ and transposing, which yields $\langle \hat A
\bsig_c(\rv')\bsig_c(\rv)\rangle^{\sf T} = -\langle\hat
\Sv_A(\rv')\beta\hat\Fv_U(\rv) \rangle^{\sf T}+\nabla' \langle \hat A
\hat\rho(\rv') \beta \hat\Fv_U(\rv)\rangle^{\sf T}$.  Averaging the
right hand side of the commutator relationship
\eqref{EQsigmacCommutator} is straightforward and requires the
averages $\langle \hat A \bsig_c(\rv')\rangle = \langle \hat A \beta
\hat\Fv_U(\rv')\rangle$ and $\langle \hat A \bsig_c(\rv) \rangle =
\langle \hat A \beta \hat\Fv_U(\rv)\rangle$.

Collecting all terms we obtain as a result the following equilibrium
hyperforce sum rule:
\begin{align}
  &-\langle\hat \Sv_A(\rv)\beta\hat\Fv_U(\rv')\rangle
  +\nabla \langle \hat A \rho(\rv) \beta \hat\Fv_U(\rv')\rangle
  \notag\\
  &+\langle\hat \Sv_A(\rv')\beta\hat\Fv_U(\rv) \rangle^{\sf T}
  -\nabla'\langle \hat A \rho(\rv') \beta \hat\Fv_U(\rv)\rangle^{\sf T}
  \notag\\
  &\quad=
    \langle\hat A\beta\hat\Fv_U(\rv')\rangle[\nabla\delta(\rv-\rv')]
    +[\nabla\delta(\rv-\rv')] \langle \hat A \beta\hat\Fv_U(\rv)\rangle.
  \label{EQsigmacCommutatorSumRule}
\end{align}
The right hand side of \eqr{EQsigmacCommutatorSumRule} can upon using
\eqr{EQsigmacAdjointAverage4} alternatively be expressed as
$[\langle\hat A\nabla'\hat\rho(\rv')\rangle - \Sv_A(\rv')]
    [\nabla\delta(\rv-\rv')] +[\nabla\delta(\rv-\rv')] [\langle\hat
      A\nabla\hat\rho(\rv)\rangle - \Sv_A(\rv)]$.

As an alternative starting point we use the adjoint of the commutator
relation \eqref{EQsigmacCommutator}, which is given by
\begin{align}
    -[\bsig_c^\dagger(\rv), \bsig_c^\dagger(\rv')] &=
    \bsig_c^\dagger(\rv')[\nabla\delta(\rv-\rv')]
    \notag\\&\quad
    +[\nabla\delta(\rv-\rv')] \bsig_c^\dagger(\rv),
    \label{EQsigmacCommutatorAdjoint}
\end{align}
Multiplying \eqr{EQsigmacCommutatorAdjoint} by the
configuration-dependent observable $\hat A(\rv^N)$ from the left and
building the thermal average on both sides involves the following
three terms:
$-\langle \hat A \bsig_c^\dagger(\rv) \bsig_c^\dagger(\rv')\rangle =
\langle [\bsig_c(\rv) \hat A][\bsig_c(\rv')-\nabla'\hat\rho(\rv')]
\rangle = \langle \hat\Sv_A(\rv)[\beta \hat\Fv_U(\rv')
  -\nabla'\hat\rho(\rv')]\rangle = \langle
\hat\Sv_A(\rv)\beta\hat\Fv_c(\rv')\rangle$, and similarly
$-\langle \hat A \bsig_c^\dagger(\rv')
\bsig_c^\dagger(\rv)\rangle^{\sf T} = \langle \beta\hat \Fv_c(\rv)
\hat\Sv_A(\rv')\rangle$, as well as $ \langle \hat A
\bsig_c^\dagger(\rv) \rangle = \langle [\bsig_c(\rv)\hat A]\rangle =
\langle \hat \Sv_A(\rv) \rangle=\Sv_A(\rv)$.
Collecting these results leads to the following sum rule:
\begin{align}
  &\langle\hat \Sv_A(\rv)\beta\hat\Fv_c(\rv')\rangle
  -\langle\beta\hat\Fv_c(\rv)\hat\Sv_A(\rv')\rangle
  \notag\\
  &=\Sv_A(\rv')[\nabla\delta(\rv-\rv')]
    +[\nabla\delta(\rv-\rv')] \Sv_A(\rv),
    \label{EQhyperforceSumRuleConfigurationalEquilibrium}
\end{align}
which is the analogous position-dependent version of the more general
(momentum-dependent) result of Ref.~\cite{mueller2024gauge}, which
features the momentum-dependent force operator $\hat\Fv(\rv)$, as
given by \eqref{EQforceDensityOperator}, in lieu of its
configurational counterpart $\hat\Fv_c(\rv)$, as given by
\eqr{EQconfigurationalForceDensityOperator}.

The examples \eqref{EQsigmacAdjointAverage2},
\eqref{EQsigmacCommutatorSumRule} and
\eqref{EQhyperforceSumRuleConfigurationalEquilibrium} demonstrate that
an algebraic structure of correlation functions emerges from both the
adjoint property \eqref{EQsigmacAdjoint} and the commutator
\eqref{EQsigmacCommutator}. We have restricted ourselves to
considering products of two shifting operators. The present shifting
in particle positions does not yet constitute a canonical transform
and we next demonstrate that incorporating the momentum transform
allows one to obtain canonical shifting transformations which retain
the useful commutator structure revealed above for purely
configurational shifting.

\begin{figure*}[!t]
  \vspace{1mm}
  \includegraphics[width=.8\textwidth]{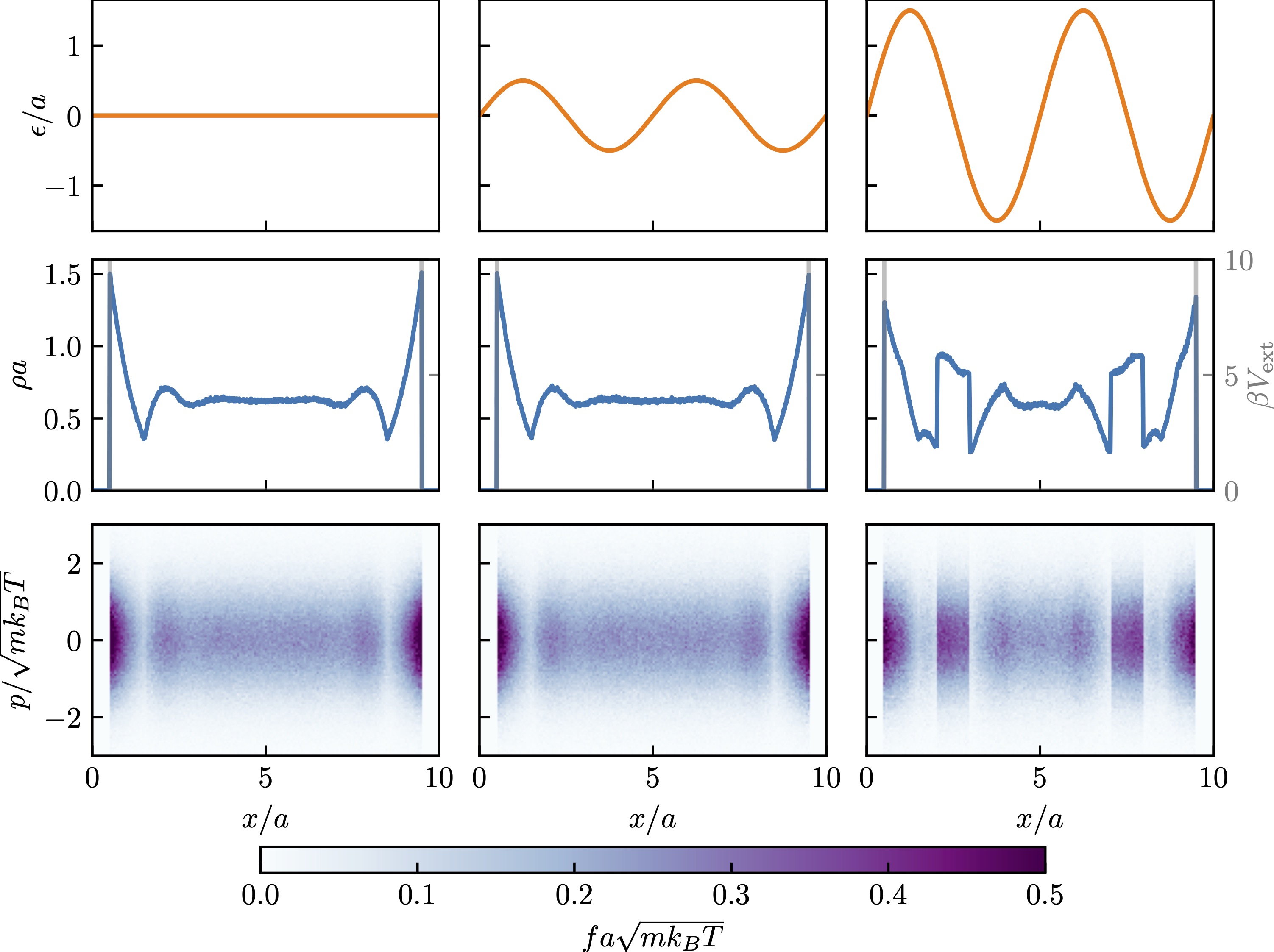}
  \caption{Demonstration of particle gauge invariance via Monte Carlo
    simulations. Hard rod particles of size $a$ are confined between
    two hard walls, as represented by a scaled external potential
    $\beta V_\rmext(\rv)$ (gray lines) that vanishes for $0.5a \leq x
    \leq 9.5a$ and is~$\infty$ otherwise.  Results are shown for the
    unshifted system (left column), a shifted system (middle column),
    and a corrupted system (right column). In the latter case the
    amplitude of the shifting field beyond the threshold of
    invertibility, such that the transformation fails to be bijective
    and is hence no longer a gauge transformation. Shown is the scaled
    sinusoidal shifting field $\epsilon(x)/a$ as a function of scaled
    position $x/a$ (top panels), the scaled density profile $\rho(x)a$
    (middle panels), and the scaled one-body phase space distribution
    $f(x,p)a \sqrt{mk_BT}$ (bottom panels), where the variables are
    the scaled position $x/a$ and scaled momentum
    $p/\sqrt{mk_BT}$. The shifting transformation displaces both
    coordinates and momenta. Despite the difference in Monte Carlo
    simulation moves and acceptance events, the results in the
    original system (left column) and in the transformed sytem (middle
    column) are numerically identical to each other, while the
    invalidly transformed system develops artifacts (right column).}
\label{FIGsimulationMC}
\end{figure*}

\subsection{Infinitesimal phase space shifts and Lie algebra}
\label{SECphaseSpaceShiftingSumRules}

We are now in a position to describe the consequence of the full phase
space shifting transformation \cite{mueller2024gauge}, as represented
on the level of microstates by the maps that displace particle
positions and momentum degrees of freedom according to
Eqs.~\eqref{EQriTransform} and \eqref{EQpiTransform},
respectively. Corresponding localized shifting operators displace in
position in analogy to the configurational shifting operators
$\bsig_c(\rv)$, as defined in \eqr{EQsigmacDefinition}.

Specifically, the full position-resolved phase space shifting
operators are defined as \cite{mueller2024gauge}:
\begin{align}
  \bsig(\rv) &= 
  \sum_i \big[
    \delta(\rv-\rv_i)\nabla_i 
    +\pv_i\nabla\delta(\rv-\rv_i)\cdot\nabla_{\pv_i}
    \big].
  \label{LQsigma}
\end{align}
When applied to a purely configuration-dependent observable $\hat
A(\rv^N)$, the dependence on momentum vanishes and $\bsig(\rv)\hat
A(\rv^N)= \bsig_c(\rv)\hat A(\rv^N)$, where $\bsig_c(\rv)$ is defined
via \eqr{EQsigmacDefinition}. The momentum contribution to the
transform [second term in the sum on the right hand side of
  \eqr{LQsigma}] constitutes spatially localized momentum shifting
operators with the explicit form:
\begin{align}
  \bsig_p(\rv) &= 
  \sum_i \pv_i\nabla\delta(\rv-\rv_i)\cdot\nabla_{\pv_i}.
\end{align}

Together with the configurational shifting operator $\bsig_c(\rv)$, we
can express $\bsig(\rv)$, as given via \eqr{LQsigma}, as the sum of
configurational and momentum contributions according to:
\begin{align}
  \bsig(\rv) &= \bsig_c(\rv) + \bsig_p(\rv).
  \label{EQsigmaAsSumConfigurationAndMomentum}
\end{align}
The adjoint momentum shift operator is given as:
\begin{align}
  \bsig_p^\dagger(\rv) &= -\bsig_p(\rv) - \nabla\hat\rho(\rv),
  \label{EQsigmapAdjoint}
\end{align}
which follows from integration by parts on momentum space, using
$\nabla\cdot\nabla_{\pvi}\pvi=\nabla\cdot\unity=\nabla$, and
identifying the density operator $\hat\rho(\rv)$, see its definition
\eqref{EQdensityOperator}.

The adjoint of the full phase space shifting operator $\bsig(\rv)$ is
then the sum of the individual adjoint identities
\eqref{EQsigmacAdjoint} and \eqref{EQsigmapAdjoint}. The result is the
simple anti-self adjoint property:
\begin{align}
    \bsig^\dagger(\rv)=-\bsig(\rv).
    \label{LQsigmaAntiSelfAdjoint}
\end{align}

Despite an increase in apparent structural complexity of $\bsig(\rv)$
over $\bsig_c(\rv)$, the corresponding commutator relationship remains
simple:
\begin{align}
  [\bsig(\rv),\bsig(\rv')] &= 
  \bsig(\rv')[\nabla\delta(\rv-\rv')] 
  + [\nabla\delta(\rv-\rv')]\bsig(\rv),
  \label{LQsigmaAlgebra}
\end{align}
which is identical in form to the corresponding configurational
identity \eqref{EQsigmacCommutator}.

The following two general properties are straightforward to show:
\begin{align}
  [\bsig(\rv), \bsig(\rv')]^\dagger &= -[\bsig(\rv), \bsig(\rv')],
  \label{EQsigmaCommutatorAntiSelfAdjoint}\\
  [\bsig(\rv), \bsig(\rv')] &= -[\bsig(\rv'), \bsig(\rv)]^{\sf T},
  \label{EQsigmaCommutatorAntiSymmetric}
\end{align}
where \eqr{EQsigmaCommutatorAntiSelfAdjoint} holds for two
self-adjoint operators and \eqr{EQsigmaCommutatorAntiSymmetric} is
valid for two vectorial operators.  Furthermore the Jacobi identity
holds:
\begin{align}
    [\sigma_a(\rv),[\sigma_b(\rv'),\sigma_c(\rv'')]] \quad&  \notag\\
    +[\sigma_b(\rv'),[\sigma_c(\rv''),\sigma_a(\rv)]] \notag \\
    +[\sigma_c(\rv''),[\sigma_a(\rv),\sigma_b(\rv')]] &=0.
\end{align}

Applying the localized phase space shifting operators to the
Hamiltonian and to the Boltzmann factor respectively yields:
\begin{align}
  -[\boldsymbol\sigma(\rv) H] &= \hat \Fv(\rv),
  \label{LQFhatFromSigma}
  \\
  [\bsig(\rv) \e^{-\beta H}] &= 
  \beta \hat\Fv(\rv) \e^{-\beta H},
  \label{LQsigmaToBoltzmannFactor}
\end{align}
where the total force density operator, including the divergences of
the explicit momentum-dependent kinetic stress tensor
$\nabla\cdot\hat\taub(\rv)$, is given by \eqr{EQforceDensityOperator}.
Equations \eqref{LQFhatFromSigma} and \eqref{LQsigmaToBoltzmannFactor}
mirror the corresponding relationships
\eqref{EQFUOperatorFromShifting} and
\eqref{EQFUOperatorFromBoltzmannShifting} for configurational shifting
on the basis of $\bsig_c(\rv)$ given by \eqr{EQsigmacDefinition} and
the potential force density $\hat\Fv_U(\rv)$ given by
\eqr{EQFUOperator}.

The procedure of obtaining sum rules from full phase space shifting is
similar in structure to working with configurational shifts described
in Sec.~\ref{SECconfigurationShifting}.  Application to a phase space
function $\hat A(\rv^N, \pv^N)$ with full configurational dependence
on $\rv^N$ and momentum dependence on $\pv^N$ yields
\cite{robitschko2024any, mueller2024gauge}:
\begin{align}
  \hat \Sv_A(\rv) 
  &= [\bsig(\rv) \hat A] \\
  &= \sum_i \delta(\rv-\rv_i)(\nabla_i \hat A)
  \notag\\ & \quad
  +\nabla\cdot \sum_i \delta(\rv-\rv_i) 
  (\nabla_{\pv_i} \hat A)\pv_i.
  \label{EQSAOperatorDefinition}
\end{align}
Application in the form $\langle [\bsig^\dagger(\rv)\hat A ]\rangle =
\langle \hat A \bsig(\rv)\rangle$ yields, upon using the self-adjoint
property \eqref{LQsigmaAntiSelfAdjoint} and the generation of the
force density operator via application to the Boltzmann factor
\eqref{LQsigmaToBoltzmannFactor}, the one-body hyperforce sum rule:
\begin{align}
  \Sv_A(\rv) + \langle \hat A \beta \hat \Fv(\rv) \rangle &= 0.
  \label{EQhyperForceSumRuleGeneral}
\end{align}
The exact identity \eqref{EQhyperForceSumRuleGeneral} applies to
general observables $\hat A(\rv^N, \pv^N)$, with the one-body
hyperforce density being the average $\Sv_A(\rv)=\langle
\hat\Sv_A(\rv)\rangle$ of the corresponding hyperforce phase space
function \eqref{EQSAOperatorDefinition}, and the full force density
operator $\hat \Fv(\rv)$ being given by
\eqr{EQforceDensityOperator}. Again we point out the formal similarity
of Eqs.~\eqref{EQSAOperatorDefinition} and
\eqref{EQhyperForceSumRuleGeneral} with the corresponding
configuration versions \eqref{EQsigmacAdjointAverage3} and
\eqref{EQsigmacAdjointAverage4}.

Building the thermal average of the commutator relationship
\eqref{LQsigmaAlgebra} yields the following hyperforce sum rule:
\begin{align}
  &    \langle \hat \Sv_A(\rv) \beta \hat \Fv(\rv') \rangle
  -\langle \beta \hat \Fv(\rv) \hat \Sv_A(\rv') \rangle
  \notag \\ & \qquad
  = \Sv_A(\rv') [\nabla\delta(\rv-\rv')]
   + [\nabla\delta(\rv-\rv')] \Sv_A(\rv),
  \label{LQsumRuleLie}
\end{align}
which is in analogy to the configurational version
\eqref{EQhyperforceSumRuleConfigurationalEquilibrium}.  The right hand
side of \eqr{LQsumRuleLie} vanishes for $\rv\neq\rv'$ and the
following exchange symmetry holds for the case of distinct positions:
\begin{align}
  \langle \hat \Sv_A(\rv) \hat\Fv(\rv') \rangle
  &= \langle \hat \Fv(\rv) \hat \Sv_A(\rv') \rangle.
  \label{LQhyperForceExchangeSymmetry}
\end{align}

It remains to formulate the Lie algebra for phase space shifting.  For
prescribed form of the shifting field $\eps(\rv)$ one uses the
localized shifting operators \eqref{LQsigma} to define a global
shifting operator via integration:
\begin{align}
  \Sigma[\eps] &= \int d\rv \eps(\rv) \cdot \bsig(\rv)
  \label{EQSigmaFromIntegration}
  \\
  &= \sum_i\big\{
    \eps(\rv_i)\cdot\nabla_i - 
    [\nabla_i\eps(\rv_i)]:\pv_i\nabla_{\pv_i}\big\}.
    \label{LQSigmaOperator}
\end{align}
where the explicit form \eqref{LQSigmaOperator} is obtained from using
$\bsig(\rv)$ according to \eqr{LQsigma} and carrying out the position
integral over $\rv$.  The colon in \eqr{LQSigmaOperator} indicates a
double tensor contraction, which is equivalently the trace of the
product of the two matrices. The phase space shifting operator
$\Sigma[\eps]$ depends functionally on the shifting field $\eps(\rv)$
as is indicated by the brackets.

\begin{figure*}[!t]
  \vspace{1mm}
  \includegraphics[width=.9\textwidth]{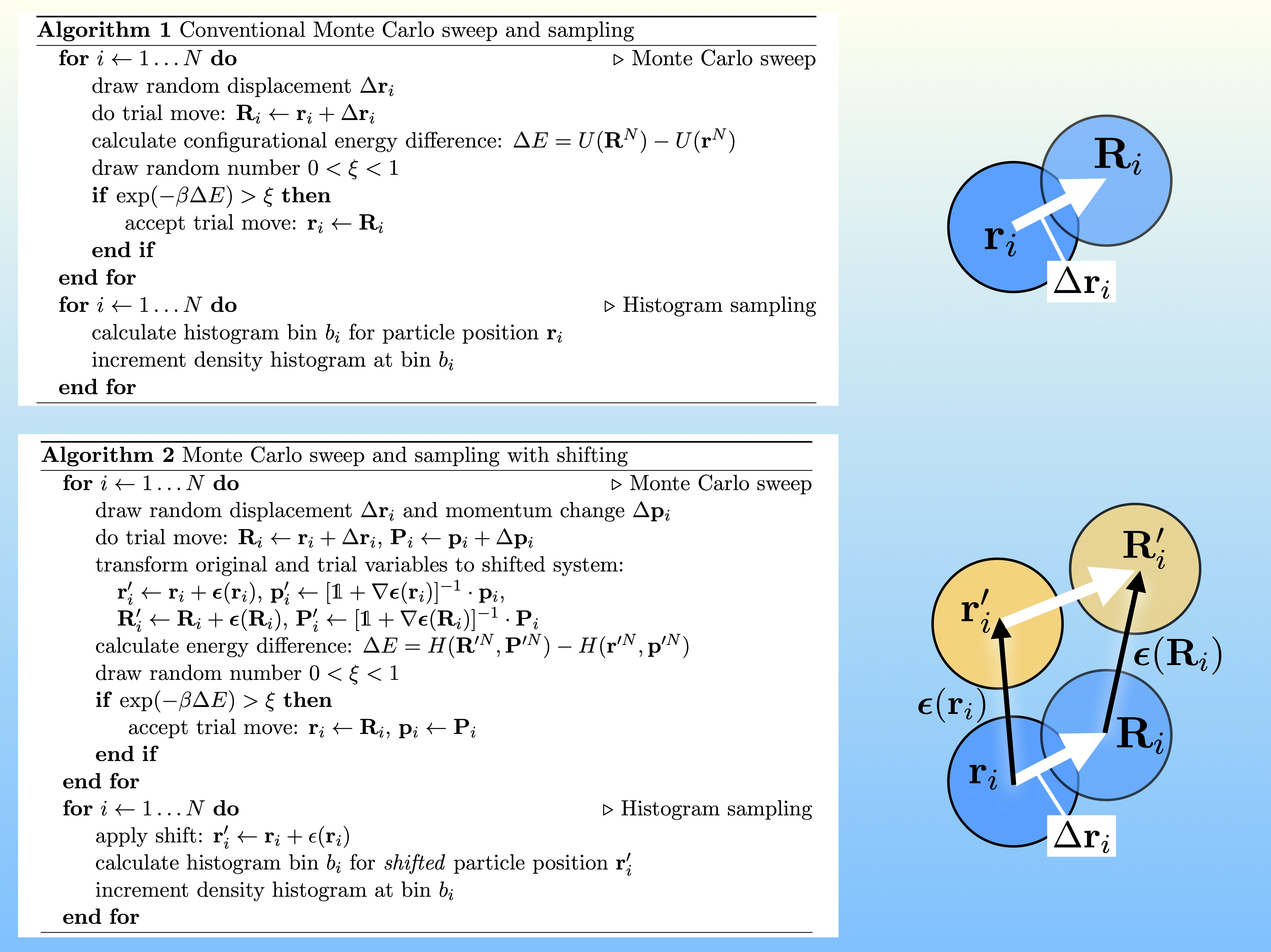}
  \caption{Pseudocodes that describe Monte Carlo simulations in the
    standard form (Algorithm~1) and using phase space shifting
    (Algorithm~2). The standard Metropolis algorithm is based on trial
    displacement from $\rvi, \pv_i$ to $\Rv_i, \Pv_i$, where the
    capitalization indicates the trial variables. In the shifted
    system the transformation is applied both to the current state,
    resulting in $\rv'_i, \pv'_i$, as well as to the trial state,
    resulting in $\Rv_i', \Pv_i'$. The Metropolis criterion is then
    evaluated with respect to the shifted variables and these then
    also form the basis for the sampling of observables.  The
    illustrations of the shifted system only display configurational
    degrees of freedom with the momentum variables being treated
    accordingly. The pseudocode for histogram sampling in the shifted
    system only describes the configurational strategy with
    corresponding momentum sampling being required for obtaining,
    e.g., the one-body phase space distribution function $f(\rv,\pv)$
    shown in Fig.~\ref{FIGsimulationMC}.}
\label{FIGpseudoCode}
\end{figure*}

The commutator relationship for full phase space shifting is given by
\begin{align}
  [\Sigma[\eps_1], \Sigma[\eps_2]] &= \Sigma[\eps_\Delta].
  \label{LQSigmaCommutator}
\end{align}
The difference shifting vector field $\eps_\Delta(\rv_i)$ is thereby
obtained from the given forms of $\eps_1(\rv_i)$ and $\eps_2(\rv_i)$
via the relation \eqref{EQepsDelta}, as previously obtained for pure
position shifting, which we reproduce for convenience:
$\eps_\Delta(\rv_i) = \eps_1(\rv_i)\cdot[\nabla_i\eps_2(\rv_i)]
-\eps_2(\rv_i)\cdot[\nabla_i\eps_1(\rv_i)]$.

The commutator relationship \eqref{LQSigmaCommutator} forms a
non-commutative Lie algebra due to the following three features: i)
anti-symmetry, which originates from the very definition of the
commutator, ii) bilinearity, which is induced by the linearity of the
differential operator \eqref{LQSigmaOperator}, and iii) the Jacobi
identity: $[\Sigma_1, [\Sigma_2,\Sigma_3]] +
[\Sigma_2,[\Sigma_3,\Sigma_1]] + [\Sigma_3, [\Sigma_1,\Sigma_2]]=0$,
which can be shown by direct manipulation using
Eq.~\eqref{LQSigmaOperator}. The functional arguments are suppressed
in the notation such that $\Sigma_1=\Sigma[\eps_1]$,
$\Sigma_2=\Sigma[\eps_2]$, and $\Sigma_3=\Sigma[\eps_3]$.

By construction, the shifting operators perform the following phase
space displacement to lowest order in the shifting field and its
gradient:
\begin{align}
  f(\rv'^N, \pv'^N) 
  &= f(\rv^N,\pv^N) + \Sigma[\eps]f(\rv^N,\pv^N).
  \label{LQfShiftedBySigma}
\end{align}
We recall the relationship of original phase space variables $\rv^N,
\pv^N$ and their displaced versions $\rv'^N$ and $\pv'^N$ via the
joint particle-resolved transformation \eqref{EQriTransform}
and~\eqref{EQpiTransform}.

For completeness, we can restore the localized shifting operators
$\bsig(\rv)$ from $\Sigma[\eps]$ by functional differentiation with
respect to the shifting field according to:
\begin{align}
  \bsig(\rv) &= \frac{\delta \Sigma[\eps]}{\delta\eps(\rv)}.
  \label{EQsigmaFromDifferentiation}
\end{align}
As a consequence, the following relationships hold between single and
successive application of localized shifting and functional
derivatives with respect to the shifting field:
\begin{align}
  \bsig(\rv) f(\rv^N,\pv^N)
  &=  \frac{\delta f(\rv'^N, \pv'^N)}{\delta\eps(\rv)} 
  \Big|_{\eps=0},
  \label{LQsigmaAsFunctionalDerivative}\\
  \bsig(\rv) \bsig(\rv') f(\rv^N,\pv^N) &= 
  \frac{\delta^2 f(\rv'^N, \pv'^N)}
       {\delta\eps(\rv)\delta\eps(\rv')} \Big|_{\eps=0}
  \notag\\&\quad
  +[\nabla\delta(\rv-\rv')] \bsig(\rv) f(\rv^N, \pv^N).
  \label{LQdoublesigmaAsFunctionalDerivative}
\end{align}

We have hence demonstrated that much of the favourable structure of
coordinate shifting, as described in
Sec.~\ref{SECconfigurationShifting}, is not only retained, see
e.g.\ the corresponding operator algebra for phase space shifting
\eqref{LQsigmaAlgebra} and coordinate shifting
\eqref{EQsigmacCommutator}, but even formally simplified, see the
simple self-adjoint nature of phase space shifting
\eqref{LQsigmaAntiSelfAdjoint}, as compared to the (slightly) more
complex version \eqref{EQsigmacAdjoint} for coordinate shifting.

The mathematical structure that is encoded in the behaviour of the
differential transformations allows one to derive exact sum rules, as
we have demonstrated. That the gauge invariance against phase space
shifting is not a mere formal device but an intrinsic property of the
statistical mechanics of many-body systems is arguably most strikingly
demonstrated for finite shifting, as we turn to next.

\section{Computer Simulations in the shifted phase space}
\label{SECsimulations}

Finite phase space shifting is expressed by the general coordinate and
momentum map \eqref{EQriTransform} and \eqref{EQpiTransform} described
in Sec.~\ref{SECfinitePhaseSpaceShifting}. We wish to demonstrate that
phase space averages are {\it in practice} invariant under this gauge
transformation. As a prototypical example we consider hard rod
particles of size $a$ in one spatial dimension. The hard core
interaction assigns vanishing statistical weight to any configuration
with two particles $i$ and $j$ having spatial distance $|x_i-x_j|<a$,
where $x_i$ and $x_j$ are one-dimensional position coordinates. We
refer the reader to Ref.~\cite{sammueller2023whyNeural} for a
description of the emerging collective physics as described on the
basis of classical density functional theory and using neural excess
free energy functionals. We deem the hard rod model to constitute a
fundamental test case. The results presented in
Ref.~\cite{mueller2024gauge} for soft interparticle and external
interaction potentials demonstrate the generality of our conclusions.

We use Monte Carlo simulations and choose two representative
observables. One is the density ``operator'', $\hat A=\hat\rho(x) =
\sum_i\delta(x-x_i)$ where $x$ denotes a generic position and we have
cast the general definition \eqref{EQdensityOperator} of
$\hat\rho(\rv)$ in one-dimensional form. We furthermore address the
one-body phase space distribution function $f(x,p)=\langle \hat
f(x,p)\rangle$, where the observable under consideration is $\hat
A=\hat f(x,p)=\sum_i\delta(x-x_i)\delta(p-p_i)$, with $p$ being a
generic momentum variable.

To carry out the simulations in the displaced system we use a
one-dimensional shifting field given by the following specific form:
\begin{align}
  \epsilon(x) = \epsilon_0 \sin(4\pi x/ L),
\end{align}
where $\epsilon_0$ is an amplitude that governs the overall magnitude
of the displacement, $L$ denotes the system size, which we choose as
$L=10a$. The system is confined between two hard walls. Setting
$\epsilon_0=0$ leads to vanishing shift and constitutes our
baseline. We compare the corresponding results with those obtained for
two different chosen values of $\epsilon_0$. We first set
$\epsilon_0/a=0.5$, which is a safe choice below the threshold of
invertibility and the resulting phase space transformation is
bijective. As a further choice we take $\epsilon_0/a = 1.5$, which is
above the invertibility threshold, which we obtain in the present
simple geometry as $\epsilon_{0}=L/(4\pi)=0.796$ from the requirement
$d(x+\epsilon(x))/dx > 0$.

The simulation results shown in Fig.~\ref{FIGsimulationMC} indicate
that sampling in the validly displaced system $\epsilon_0/a=0.5$
(middle column) generates results that are numerically identical to
those of the original system with $\epsilon_0/a=0$ (left column). The
invalid transformation, upon using identical code and merely changing
the amplitude of the shifting field to the value of
$\epsilon_0/a=1.5$, yields corrupted results (right column). Such
behaviour is expected for this case of a transform that is not
bijective.

Pseudocode for the corresponding Metropolis Monte Carlo algorithm
based on single particle moves is described in
Fig.~\ref{FIGpseudoCode}. We lay out the standard procedure as a
reference (algorithm~1), together with the changes that implement
phase space shifting both for the generation of the microstates and
for the data acquisition via histogram filling (algorithm~2).

\section{Conclusions}
\label{SECconclusions}

In conclusion we have described background and details for the gauge
transformation of statistical mechanical microstates of
Ref.~\cite{mueller2024gauge}. Thereby the thermal invariance against
phase space shifting implies exact sum rules, as previously identified
in a variety of different settings \cite{hermann2021noether,
  hermann2022topicalReview, hermann2022variance, hermann2022quantum,
  sammueller2023whatIsLiquid, hermann2023whatIsLiquid,
  robitschko2024any, tschopp2022forceDFT}. These include inhomogeneous
thermal quantum systems \cite{hermann2022quantum}, as well as
classical liquids and more general soft matter states
\cite{sammueller2023whatIsLiquid, hermann2023whatIsLiquid}. Previous
work was based on variational methods and using invariance against the
specific form of the displacement field $\eps(\rv)$ that parameterizes
the phase space transformation.  Here we have given background and
details about the structure of the underlying invariance group
including the algebra of phase space differential operators identified
in Ref.~\cite{mueller2024gauge}. The relevant phase space operators
represent infinitesimal versions of finite gauge transformations and
the analysis of their properties allows one to reveal rich
mathematical structure.

The gauge transformation affects particle position
\eqref{EQriTransform} and momentum \eqref{EQpiTransform} degrees of
freedom. The vector field $\eps(\rv)$ that parameterizes the transform
needs to be smooth and of a form that allows the transformation to be
bijective. Specifically we require that the function
$\rv_i+\eps(\rv_i)$ is a diffeomorphism.  With the position transform
being very general, the momentum transform ensures that the
differential phase space volume element is conserved and that the
joint transformation \eqref{EQriTransform} and \eqref{EQpiTransform}
is canonical in the sense of classical mechanics \cite{goldstein2002}.
This is not the case in an early contribution by Baus and Lovett
\cite{baus1992}, who have worked with functional differentiation
methods based on only the position transformation
\eqref{EQriTransform}, rather than the present phase space operators
and their algebraic commutator structure.

For finite displacement we have described in detail the geometric
nature of position shifting. This transformation can be inverted and
chained and hence establishes a group structure. Notwithstanding the
apparent geometric simplicity, the group is non-commutative with the
order of two consecutive shifts being relevant.  Despite its more
complex matrix multiplication structure, the corresponding momentum
transform \eqref{EQpiTransform} remains entirely compatible with the
mathematical structure of the position transform.

When considering the infinitesimal version of shifting much additional
structure is revealed by formalizing relationships for differential
operators that perform the shifting.  The transformations are
functionally dependent on the vector field $\eps(\rv)$ that
parameterizes the transform. Functional differentiation yields
corresponding commutator relationships for position-localized
differential operators that act on the phase space of a many-body
system.  We have laid out the properties of pure position shifting and
have compared against those of full phase space gauge
transformation. Both versions of corresponding localized differential
operators can be used as a basis for deriving exact statistical
mechanical sum rules, as we have demonstrated.  The mathematical
derivation of these correlation function identities (sum rules) is
greatly simplified by the use of the differential operator formalism.

The classical gauge theory that we have presented carries strong
similarities with the theory of Lie groups and associated Lie
algebras. In particular, the phase space shifting operators
$\Sigma[\eps]$ given by \eqr{LQSigmaOperator} form a non-commutative
Lie algebra, as described by the commutator relation
\eqref{LQSigmaCommutator}.  For pure configuration shifting the
corresponding Lie algebra is given by
\eqr{EQshiftingCommutatorViaLieBracket}. The dependence on the
shifting field $\eps(\rv)$ remains thereby explicit and it acts to
identify the elements of the algebra.

In the present statistical mechanical setting it is useful to go
further and eliminate the dependence on $\eps(\rv)$. We recall that
$\eps(\rv)$ possesses the status of a mere gauge function, which does
not affect the actual physics; see the similarities with gauge
invariance in electrodynamics as described in
Sec.~\ref{SECelectrodynamics}.  Functional differentiation of the Lie
algebra with respect to the shifting field constitutes a
well-characterized route to eliminate the dependence on $\eps(\rv)$,
which in turn generates spatial localization via Dirac distributions.
The resulting operator commutator identities, see \eqr{LQsigmaAlgebra}
for full phase space shifting and \eqr{EQlieAlgebraQ} for the purely
configurational version, retain much of the favourable structure of a
Lie algebra, see the occurrence of structure constants
\eqr{EQstructureConstantsPositionShifting}. However, the distribution
character reaches beyond elementary Lie theory \cite{robbin2022book}
and it would be interesting in future work to explore connections with
more abstract Lie concepts in modern mathematical treatments of the
subject.

We have described several implications of the gauge invariance for
simulation methods focusing on Monte Carlo sampling. Choosing a
specific form of the shifting field and displacing the particles in a
prototypical confined one-dimensional hard core system allows one to
give a standalone demonstration of gauge invariance.  Applying the
transformation \eqref{EQriTransform} and \eqref{EQpiTransform} goes in
practice beyond a mere change of phase space variables, as this
procedure genuinely alters the resulting Markov chain due to the
evaluation of acceptance probabilities in the virtually displaced
system. Nevertheless, the invariance property ensures that thermal
averages remain valid despite the seemingly violated requirement of
detailed balance.  That the acceptance rates differ in the displaced
and original systems serves as a practical indicator for the
algorithmic differences. The sampled states retain, as theoretically
predicted, identical one-body density profile and phase space
distribution function, which we took as representative of the
behaviour of general observables.

In future work, it would be interesting to investigate the potential
use of gauge invariance in machine learning of neural density
functionals \cite{sammueller2023neural, sammueller2023whyNeural}, for
constructing analytical density functional approximations
\cite{gul2024testParticle}, for cross fertilization with hyperdensity
functional theory~\cite{sammueller2024hyperDFT,
  sammueller2024whyhyperDFT}, for the construction of advanced
simulation schemes \cite{borgis2013, delasheras2018forceSampling,
  coles2019, coles2021, rotenberg2020, purohit2019, schultz2016,
  trokhymchuk2019, schultz2019, purohit2019} in particular based on
mapped averaging \cite{moustafa2015, schultz2016, moustafa2017jctp,
  moustafa2017prb, lin2018, moustafa2019, schultz2018, purohit2018,
  schultz2019, purohit2019, trokhymchuk2019, purohit2020,
  moustafa2022}, and for the application to mixtures
\cite{matthes2024mix}.

\acknowledgments We thank Sophie Hermann for useful discussions.  This
work is supported by the DFG (Deutsche Forschungsgemeinschaft) under
project~551294732.

\appendix

\section{Functional derivatives in electrodynamics}
\label{SECappendixElectrodynamics}

We denote the transformed action by $S_\rmext[\varphi]$, as given via
the right hand side of \eqr{EQSextTransformed}. The brackets indicate
functional dependence, which arises as the value of the action depends
a priori on the specific form of the gauge function $\varphi(x)$. We
can express the gauge invariance as the identity
\begin{align}
  S_\rmext=S_\rmext[\varphi],
  \label{EQactionInvariant}
\end{align}
where the left hand side denotes the original action \eqref{EQSext},
which is clearly independent of $\varphi(x)$. The invariance
equation~\eqref{EQactionInvariant} holds irrespective of the form of
the gauge function $\varphi(x)$. Hence a valid identity is retained
upon functionally differentiating both sides by~$\varphi(x)$. The left
hand side of \eqr{EQactionInvariant} then vanishes as there is no
dependence on the gauge function. 

Differentiating also on the right hand side of \eqr{EQactionInvariant}
gives $0=\delta S_\rmext[\varphi]/\delta\varphi(x) = \delta [S_\rmext
  + \int dx' \varphi(x')\partial_\nu'J^\nu(x')]/\delta\varphi(x)$,
where the prime indicates a new spacetime integration variable and
$\partial_\nu'$ is the corresponding spacetime derivative.  The first
term in the sum vanishes, as argued above.  Exchanging in the second
term the order of the functional derivative and the spacetime integral
gives $0=\int
dx'[\delta\varphi(x')/\delta\varphi(x)]\partial_\nu'J^\nu(x')= \int
dx' \delta(x-x')\partial_\nu'J^\nu(x') = \partial_\nu J^\nu(x)$. Here
we have used that functionally differentiating a function by itself
gives the Dirac distribution, $\delta \varphi(x')/\delta
\varphi(x)=\delta(x-x')$, which is here in four dimensions. Recalling
that the result of the functional derivative vanishes, we have hence
rederived the charge continuity equation \eqref{EQcontinuity}.

While requiring slightly more steps than the route considered in the
main text, the present method via functional calculus is arguably as
generally applicable as is ordinary multivariate calculus and hence
there are no principal limits in terms of complexity of the functional
dependence under investigation. Moreover, as was demonstrated recently
for statistical functionals \cite{sammueller2023neural,
  sammueller2023whyNeural, stierle2024autodiff}, powerful software
tools that implement automatic differentiation
\cite{baydin2018autodiff} can be used.

For completeness, the free field action \eqref{EQSfree} is unaffected
by the gauge transformation, as $F_{\lambda\nu}(x)$ is already an
invariant. We demonstrate this explicitly as follows. The free field
action \eqref{EQSfree} acquires the following apparent change under
the gauge transformation \eqref{EQgaugeTransformationCovariant}
according to
\begin{align}
  S_{\rm free} & \to 
  S_{\rm free} 
  + \mu_0^{-1} \int dx 
  \varphi(x) \partial_\lambda\partial_\nu F^{\lambda\nu}(x).
\end{align}
From invariance of the free field action one can conclude that
$\partial_\lambda\partial_\nu F^{\lambda\nu}(x)=0$, which already
follows in an elementary way from the antisymmetry
$F^{\lambda\nu}(x)=-F^{\nu\lambda}(x)$.

Again an alternative derivation can be based on functional
differentiation which, following steps analogous to those described
above for the external action, gives the same result: $\delta S_{\rm
  free}[\varphi]/\delta \varphi(x)=\partial_\lambda\partial_\nu
F^{\lambda\nu}(x)/\mu_0=0$.

To make the covariant formulation more explicit, the relationship of
$F_{\nu\lambda}(x)$ to the electrical field and to the magnetic
induction is via the explicit components: $F_{0\lambda}(x)=(0, E_x/c,
E_y/c, E_z/c)$, $F_{1\lambda}(x)=(-E_x/c,0,-B_z,B_y)$,
$F_{2\lambda}(x)=(-E_y/c,B_z,0,-B_x)$
$F_{3\lambda}(x)=(-E_z/c,-B_y,B_x,0)$, where ${\bf E}(\rv,t)=(E_x,
E_y, E_z)$ and ${\bf B}(\rv,t)=(B_x, B_y, B_z)$. For completeness, the
entire field tensor reads as follows:
\begin{align}
  F_{\nu\lambda}(x)&=
  \left({\begin{array}{cccc}
    0 & E_x/c & E_y/c & E_z/c\\
    -E_x/c & 0 & -B_z & B_y\\
    -E_y/c & B_z & 0 & -B_x \\
    -E_z/c & -B_y & B_x & 0\\
  \end{array} } \right).
\end{align}


\end{document}